\RequirePackage{fix-cm}
\documentclass[smallextended]{svjour3}       % onecolumn (second format)
\smartqed  % flush right qed marks, e.g. at end of proof
\usepackage{graphicx}

\usepackage{amsmath}
\usepackage{amssymb}
\usepackage{subfigure}
\usepackage{epstopdf}
\usepackage{multirow}
\usepackage{url}
\usepackage{array}

\newcommand{\tabincell}[2]{\begin{tabular}{@{}#1@{}}#2\end{tabular}}

\begin{document}

\title{Utilizing Embeddings for Ad-hoc Retrieval by Document-to-document Similarity%\thanks{Grants or other notes
}
\subtitle{}


\author{Chenhao Yang$^1$         \and
        Ben He$^2$ \and
        Yanhua Ran$^3$
}


\institute{School of Computer \& Control Engineering \\
			University of Chinese Academy of Sciences$^{123}$ \\
              \email{benhe@ucas.ac.cn}$^2$           %  \\
}


\maketitle

\begin{abstract}
Latent semantic representations of words or paragraphs, namely the embeddings, have been widely applied to information retrieval (IR). One of the common approaches of utilizing embeddings for IR is to estimate the document-to-query (D2Q) similarity in their embeddings. As words with similar syntactic usage are usually very close to each other in the embeddings space, although they are not semantically similar, the D2Q similarity approach may suffer from the problem of ``multiple degrees of similarity''. To this end, this paper proposes a novel approach that estimates a semantic relevance score (SEM) based on document-to-document (D2D) similarity of embeddings. As Word or Para2Vec generates embeddings by the context of words/paragraphs, the D2D similarity approach turns the task of document ranking into the estimation of similarity between content within different documents. Experimental results on standard TREC test collections show that our proposed approach outperforms strong baselines.
\keywords{Information retrieval \and Word or paragraph embeddings \and Word2vec}
\end{abstract}

\section{Introduction} \label{sec:introduction}

Despite the effectiveness of the existing IR models, such as BM25 \cite{robertson1996okapi} and language modeling approach to IR \cite{Ponte1998}, they can be potentially further improved by utilizing the latent semantic representations of words and documents.
According to the recent advances in natural language processing (NLP), words and texts can be represented with semantically distributed real-valued vectors, i.e. the \textit{embeddings}, generated by neural network models \cite{Mikolov2013a,DBLP:journals/corr/LeM14}. The word and text embeddings have been shown to be effective and efficient in many NLP tasks \cite{Mikolov2013a,DBLP:journals/corr/LeM14}. 

One of the popular approaches for utilizing embeddings for IR is to estimate the document-to-query (D2Q) similarity in their embeddings. For example, a previous study \cite{Vuli2015Monolingual} suggests that the IR effectiveness can be potentially improved by considering the semantic similarity between a document and the given query. As pointed out in \cite{Mikolov2013Linguistic}, not only semantically similar words tend to be close to each other in the embeddings space, but also words that can have ``multiple degrees of similarity''. For example, the embeddings of ``summer Olympics'' is likely to be very similar to ``winter Olympics'' because ``summer'' and ``winter'' have similar syntactic usage, and are very close to each other in the embeddings space. However, the two phrases are not semantically similar, hence the problem of ``multiple degrees of similarity''.

In this paper, we aim to improve the retrieval performance by utilizing the embeddings. The major contribution of this paper is the proposal a novel D2D similarity approach. In Section \ref{subsec:D2D_example}, we show that by estimating the similarity between a given document and a highly relevant document, the task of document ranking is turned into the similarity estimation between content within different documents. Since embeddings techniques like Word2Vec generates embeddings by predicting words within a given local context, our D2D similarity approach better utilizes the content information, and is expected to result in improved retrieval performance. In our approach, the highly relevant document is simulated by the top-ranked documents, namely the pseudo relevance document set. To the best our knowledge, this paper is the first to utilize the embeddings to estimate a semantic relevance score by the similarity between a document and the pseudo feedback set. 

Continue with the example above, for a query like ``Summer Olympics'', using our approach, the semantic relevance score is measured by the semantic similarity of a document and a pseudo relevance set, in which the documents are likely to be about summer Olympics sports such as swimming and basketball. These sports appear in different context of winter Olympics sports such as curling. In this case, a document about winter Olympics is likely to receive a low similarity to the pseudo feedback set, such that the problem of ``multiple degrees of similarity" can be overcome. The effectiveness of our approach may depend on the quality of the feedback set, and the pseudo feedback documents are not always relevant. However, as shown by the IR literature, pseudo relevance feedback (PRF) is in general useful for top-k retrieval \cite{Gonzalez2007jasist}, hence the potential improvement brought by our approach over the baselines. Indeed, our study in Section \ref{sec:quality_study} shows that the quality of the pseudo feedback set does not correlate with the effectiveness of our D2D similarity approach. In addition, the experimental results show that our approach outscores a number of recently proposed methods that attempt to utilize word/paragraph embeddings for IR.

In our study, we utilize the Word2Vec and Para2Vec techniques proposed by Mikolov et al. \cite{Mikolov2013a,DBLP:journals/corr/LeM14} to generate word and document embeddings. Word or Para2Vec is a three-layer neural net for text processing. It turns words or texts into a numerical form that can be understood by deep nets, and it has been used in several existing approaches that aim to improve IR performance by utilizing semantic relation \cite{Ai2016Improving,Rekabsaz2016Uncertainty,Vuli2015Monolingual}.
%In addition, Latent Dirichlet Allocation (LDA) \cite{blei2003latent} and TF-IDF \cite{Manning2008ir} are compared to Word or Para2Vec in generating the vector representations of documents in our experiments.

%\textcolor{blue}{To Update}

The remainder of this work is organized as follows. Section \ref{sec:related_work} introduces the related work, including classical state-of-the-art IR models (BM25 and QLM), the word and and paragraph embeddings techniques, and recent applications of embeddings to IR. Section \ref{sec:approach} compares D2D similarity with D2Q similarity and introduces our proposed D2D similarity approach. Experimental Settings and evaluation results are presented in Sections \ref{sec:exp_settings} and \ref{sec:eval_results}, respectively. In Section \ref{sec:discuss}, we analyze the sensitivities of our approach to the tunable parameters, the impact of the quality of pseudo relevance document set on our approach, compare our approach with some other recently proposed state-of-the-art approaches mainly based on embeddings, and apply our approach to Clinical Decision Support. Finally, Section \ref{sec:conclusions} concludes our work and suggests possible future research directions.

%\textcolor{red}{TODO: Add correlation between AP@k and improvement, table+figure}

%------------------------------------------------

\section{Related Work} \label{sec:related_work}

In this section, we introduce classical state-of-the-art IR models (BM25 and QLM), the word and paragraph embedding techniques and recent works that aim to improve IR effectiveness by utilizing embeddings.

%\begin{comment}
\subsection{Classical Models}
This section introduces the popular IR models that are used as baselines in this paper, including BM25 \cite{robertson1996okapi}, the query likelihood model (QLM) \cite{Ponte1998,Manning2008ir}.
As one of the most established retrieval models, BM25 computes the relevance score of a document $d$ for given query $Q$ by the following formula \cite{robertson1996okapi}:
\begin{eqnarray}
	\label{eq:bm25}
	score(d,Q)=\sum_{t \in Q}w_{t}\frac{(k_{1}+1)tf}{K+tf}\frac{(k_{3}+1)qtf}{k_{3}+qtf}
\end{eqnarray}
\noindent where $qtf$ is the frequency of query term $t$ in query $Q$. $K$ is given by $k_{1}((1-b)+b \cdot \frac{l}{avg\_l})$, in which $l$ and $avg\_l$ denote the length of document $d$ and the average length of documents in the collection, respectively. The length of document $d$ refers to the number of tokens occurring in $d$. $k_1$, $k_3$ and $b$ in Equation (\ref{eq:bm25}) are free parameters whose default setting is $k_1=1.2$, $k_3=1000$ and $b=0.75$, respectively \cite{robertson1996okapi}. $w_t$ is the IDF factor used to measure the weight of query term $t$, which is given by:
\begin{eqnarray}
	w_t=\log_{2}\frac{N-df_{t}+0.5}{df_t+0.5}
\end{eqnarray}
\noindent where $N$ is the number of documents in the collection, and $df_{t}$ is the document frequency of query term $t$ which denotes the number of documents that $t$ occurs. According to Equation (\ref{eq:bm25}), we can see that BM25 mainly consists of three variables, i.e. a TF part, an IDF part and a document length normalization part.

Ponte \& Croft propose a query likelihood model (QLM) of the language modeling approach, in which the language model of each document is estimated and then the documents are ranked by the likelihood of the query according to the estimated language model. Given a document $d$, the query likelihood $p(Q|d)$ based on the logarithmic framework is given by \cite{Ponte1998}\cite{Manning2008ir}:
\begin{eqnarray}
	\log p(Q|d)=\sum_{q \in Q} \log \frac{p(q|d)}{\alpha_{d} \cdot p(q|C)}+|Q| \cdot \log \alpha_{d}
\end{eqnarray}
\noindent where $p(q|d)$ and $p(q|C)$ denote the document language model and the collection language model, respectively. $|Q|$ is the length of query $Q$. $\alpha_{d}$ is a document-dependent constant. Relevance model \cite{lavrenko2001relevance} is one of the state-of-the-art relevance feedback methods for language modeling approach to IR. In our experiments, the popular RM3 relevance model \cite{lavrenko2001relevance} is applied to QLM, which is used as one of our baselines.

%Paik proposed a novel parameter-free TF-IDF term weighting scheme (MATF \cite{Paik2013A}), in which two term frequency normalization methods are applied to determine the importance of a query term. The MATF model is adaptive to the query length.
%\end{comment}

\subsection{Word and Paragraph Embedding Techniques} \label{sec:we_pe_technique}
The embedding techniques aim to learn low-dimensional representations for words or paragraphs. The learned embeddings are often considered to be able to capture the semantic information and are successfully applied in many NLP tasks. Topic models, such as Latent Semantic Indexing (LSI) \cite{deerwester1990indexing}, Probabilistic Latent Semantic Analysis (PLSA) \cite{hofmann1999probabilistic} and Latent Dirichlet Allocation (LDA) \cite{blei2003latent}, utilize global statistical information of the corpus. The recently emerged Word2Vec \cite{Mikolov2013a} and Paragraph Vector \cite{DBLP:journals/corr/LeM14} are based on local context information instead. Next, we introduce the skip-gram model of Word2Vec and Paragraph Vector which are used in this paper.

The Skip-gram model consists of three layers, i.e. an input layer, a projection layer and an output layer, and the objective is to predict the context of a given word $w$. 

Considering the conditional probability $p(c(w)|w)$ given a word $w$ and the corresponding context $c(w)$, the goal of Skip-gram model is to maximize the likelihood function by optimizing the parameters $\theta$ in $p(c(w)|w;\theta)$ as described below \cite{DBLP:journals/corr/GoldbergL14}:
\begin{eqnarray}
	\label{eq:skip_gram_orig}
	\mathop{\arg max}_{\theta}\prod_{(w,c(w)) \in D}p(c(w)|w;\theta)
\end{eqnarray}
\noindent where $(w,c(w))$ is a training sample and $D$ is the set of training samples. $w$ and $c(w)$ denote a word and the corresponding context, respectively. Under the assumption of independence of words in $c(w)$, Equation (\ref{eq:skip_gram_orig}) can be equivalently reformulated as follows:
\begin{eqnarray}
	\mathop{\arg max}_{\theta}\prod_{w \in Text}\left[\prod_{w' \in c(w)}p(w'|w;\theta)\right]
\end{eqnarray}
\noindent where $w'$ denotes one of the words in the context of word $w$. In addition, the conditional probability $p(c(w)|w;\theta) $ is modeled as Softmax regression which is given by:
\begin{eqnarray}
	\label{eq:softmax}
	p(c(w)|w;\theta)=\frac{e^{v_{w}\cdot v_{c(w)}} }{\sum_{c(w)' \in C}e^{v_{w}\cdot v_{c(w)'}} }
\end{eqnarray}
\noindent where $v_w$ and $v_{c(w)}$ are the n-dimensional distributed representations of word $w$ and the corresponding context $c(w)$, respectively. Substituting Equation (\ref{eq:softmax}) back into Equation (\ref{eq:skip_gram_orig}), the final objective function of Skip-gram is given by:
\begin{align}
	\label{eq:skip_gram_final}
	& \mathop{\arg max}_{\theta}\prod_{(w,c(w)) \in D}\log p(c(w)|w) \notag \\
	={} & \sum_{(w,c(w)) \in D} \left( \log e^{v_{w} \cdot v_{c(w)}}-\log\sum_{c'}e^{v_{w} \cdot v_{c(w)'}}  \right)
\end{align}
\noindent where parameters in Equation (\ref{eq:skip_gram_final}) are trained by stochastic gradient ascent method.

Paragraph Vector is inspired by the recent work in learning vector representations of words using neural networks \cite{DBLP:journals/corr/LeM14}. For a given word \textit{w}, Paragraph Vector aims to predict not only the word's context \textit{c(w)} but also a special ``word'' denoted as \textit{paragraph id}, added to represent the document. \textit{Paragraph id} is always regarded as a part of the context of each word, i.e. the actual context of word \textit{w} during the training is \textit{\{Paragraph id, c(w)\}}. \textit{Paragraph id} acts as a memory that remembers what is missing from the current context. Note that the training procedure of document embeddings in \textit{Paragraph Vector} is the same as word embeddings in \textit{Word2Vec}. At the end of training, the embedding of \textit{Paragraph id} is used to represent the document, since the embedding of \textit{paragraph id} remembers the semantic information in the document.

\subsection{Applications of the Embeddings to IR} \label{sec:embedding_ir}

Inspired by the successful applications of embeddings to NLP tasks, there have been works that attempt to utilize embeddings to enhance the retrieval performance for IR in recent years \cite{mitra2017neural} . In general, these works can be roughly divided into the following two categories.

The first category utilizes embeddings to reformulate the query. The reformulated query is expected to better reflect user's information need and can improve the retrieval performance of the baseline model. In order to solve the low matching ratio of keywords problem in sponsored search, Grbovic et al. \cite{DBLP:conf/sigir/GrbovicDRSB15} map queries into an embeddings space and expand a given query via the K-nearest neighbor queries. Kuzi et al. \cite{DBLP:conf/cikm/KuziSK16} adopt two methods to expand the query. One computes the cosine similarity between the embedding of a given term and the sum of the embeddings of all terms in a query. The other computes the cosine similarity between the embeddings of a given term and each query term, and then combines the term-to-term similarities. Roy et al. \cite{DBLP:journals/corr/RoyPMG16} propose a query expansion technique based on word embeddings in which expansion terms are obtained by K-nearest neighbor approach. The query term weighting is an important factor to influence the performance of a IR model. A good query term weighting framework can effectively improve the retrieval performance. To address the problem above, Zheng \& Callan \cite{Zheng2015Learning} propose a framework for learning term weights using word embeddings. Zamani \& Croft \cite{DBLP:conf/ictir/ZamaniC16a} propose two expansion methods to estimate accurate query language models based on word embeddings and an embedding-based relevance model. To obtain discriminative similarity scores, the cosine similarity is transformed by the sigmoid function. Roy et al. \cite{DBLP:conf/cikm/RoyGMJ16} view the embeddings of terms as points in a embeddings space and utilize kernel density estimation algorithm to estimate a relevance model. Zamani \& Croft \cite{Zamani2016Estimating} propose to estimate dense vector representations for queries based on the individual embedding vectors of vocabulary terms. The embeddings used in query reformulation can be obtained by two methods: global and local methods. The former is trained using all the documents in the collection while the latter is trained using the topic-specific documents. Diaz et al. \cite{DBLP:conf/acl/0001MC16} explore the difference between global embeddings and local embeddings when applied in query expansion. They utilize the documents returned by the initial retrieval to train local embeddings and the experiment results indicated that local embeddings provide better similarity measures than global embeddings for query expansion. 

The second category utilizes the term-term, term-document and query-document semantic similarities to improve the performance of IR models. Clinchant \& Perronnin \cite{Clinchant2013Aggregating} view a document as bag-of-embedded-words (BoEW), and then non-linearly map the word embeddings into a higher-dimensional space and aggregate them into a document-level representation. Ganguly et al. \cite{DBLP:conf/sigir/GangulyRMJ15} propose a generalized language model (GLM) which assumes that a query is sampled in the following three ways: direct term sampling, transformation via document sampling and transformation via collection sampling. The transformation probability between two terms is computed by the cosine similarity between their embeddings. The experimental results show that GLM significantly outperforms the standard LM and LDA-LM. Be similar to \cite{DBLP:conf/sigir/GangulyRMJ15}, Zuccon et al. \cite{DBLP:conf/adcs/ZucconKBA15} propose to incorporate word embeddings within the translation language framework by capturing the semantic relationships between terms. Most of existing works using Word2Vec embeddings only keep the input projections. Differently from these works, Mitra et al. \cite{DBLP:journals/corr/MitraNCC16} introduce a novel documents ranking model named DESM in which a query term score is computed by the cosine similarity between its embedding and the embedding of the document, using both the input and output projections. There are also some works which directly learn embeddings for queries and documents from the texts using deep neural network. They utilize click-through data to train a deep neural network, such as MLP or CNN, which can directly maps queries or document into embeddings. DSSM model in \cite{DBLP:conf/cikm/HuangHGDAH13} and CLSM model in \cite{DBLP:conf/cikm/ShenHGDM14} are typical examples of the above deep neural network methods. Guo et al. \cite{DBLP:conf/cikm/GuoFAC16a} view the semantic relationships between queries and documents in the perspective of transportation and the transportation gain is measured by the semantic similarity between two words when transport a document word to a query word.

There are also several approaches that do not exactly fall into any of the two categories above. Ai et al. \cite{Ai2016Improving} propose several improvements over the original Paragraph Vector model to make it more adaptive to the IR scenario. Guo et al. \cite{DBLP:conf/cikm/GuoFAC16} propose a novel relevance matching model named DRMM using deep neural network for ad-hoc retrieval. While the list of previous studies mentioned in this paper may not be exhaustive, it includes most published results on the related subject that we are aware of. Interested readers are referred to \cite{mitra2017neural} for a more complete review.

In addition to those second category of methods mentioned above, Vuli\'{c} \& Moens propose a D2Q similarity approach for for monolingual and cross-lingual IR that scores a document d for a give query Q as follows \cite{Vuli2015Monolingual}:
\begin{eqnarray}
	\label{eq:qd_doc_score}
	score(d,Q)=\lambda R(d,Q)+(1-\lambda)Sim(d,Q)
\end{eqnarray}
\noindent where $R(d, Q)$ is the score given by the baseline model, e.g. BM25 or QLM. $\lambda$ is the hyper-parameter to control the influence of $R(d, Q)$ and $Sim(d, Q)$. $Sim(d, Q)$ is the semantic similarity between $d$ and $Q$, which is measured by:
\begin{eqnarray}
	\label{eq:qd_doc_sim}
	Sim(d,Q)=\frac{\vec{d} \cdot \vec{Q}}{||\vec{d}|| \times ||\vec{Q}||}
\end{eqnarray}
\noindent where $\vec{d}$ and $\vec{Q}$ is the embedding representations of $d$ and $Q$ respectively.

As described in Section \ref{sec:introduction}, due to the problem of ``multiple degrees of similarity", the query-document semantic similarity may not be able to lead to sufficient improvement in the retrieval effectiveness. Instead, our approach measures the semantic similarity between a document and the corresponding pseudo relevance feedback set, in which the context of keywords within different documents is taken into account such that documents about the query topic can receive higher rankings.

\begin{figure}
	\centering
	\includegraphics[scale=0.35]{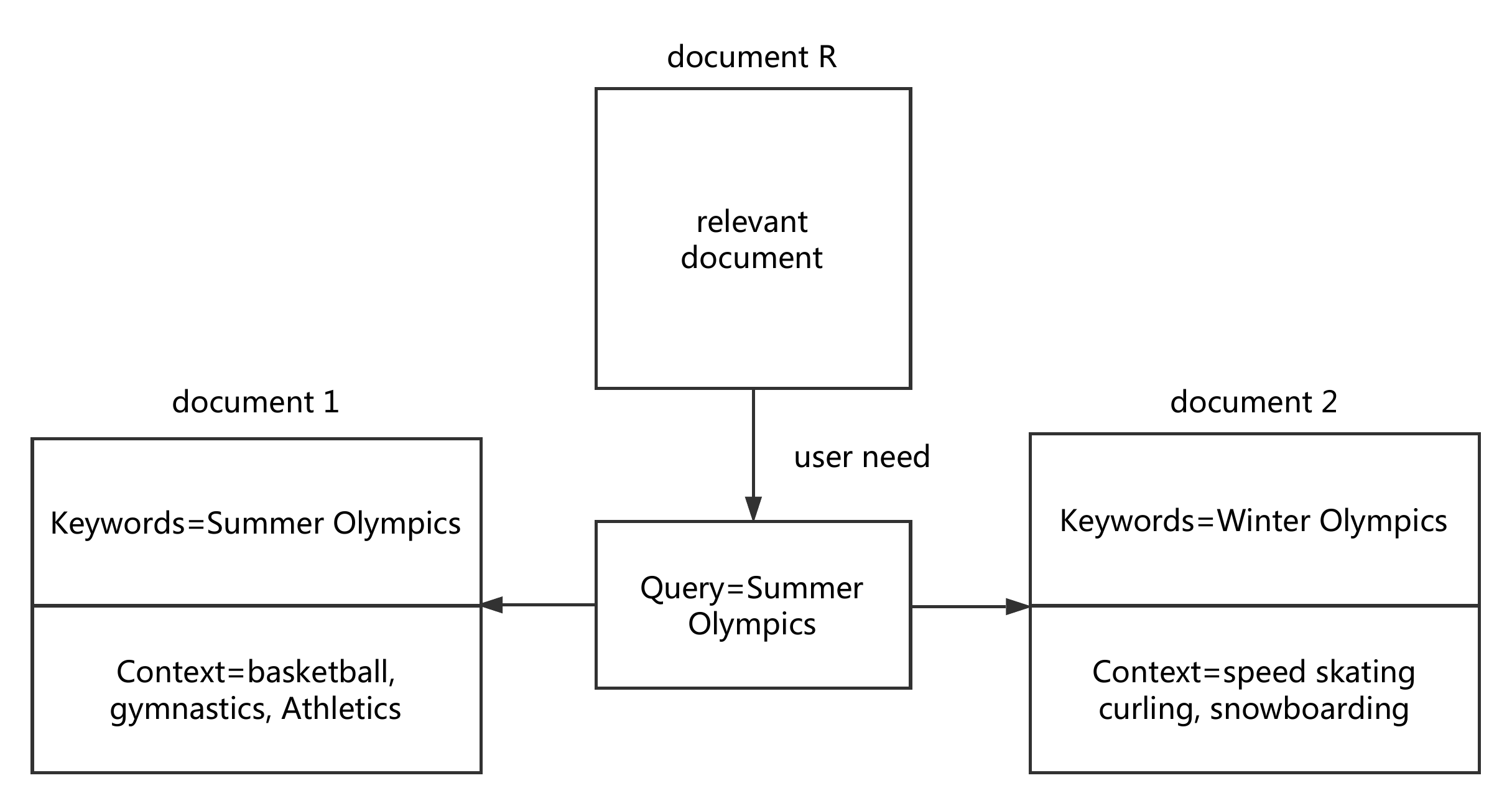}
	\caption{Example: D2D similarities. Document 1 about ``Summer Olympics'' is likely to be more relevant to the query than document 2 about ``Winter Olympics''. R is a highly relevant document. The D2Q similarity of $d_i$ is $SEM_{d\text{-}Q}(d_i, Q)=Cos(\vec{Q}, \vec{KWs_i}) + Cos(\vec{Q}, \vec{Ctx_i})$. The D2D similarity of $d_i$ to R is $SEM_{d\text{-}d}(d_i, Q) = SEM_{d\text{-}Q}(d_i, Q) + Cos(\vec{Ctx_R},\vec{Ctx_i})$. Document ranking is determined by the cosine similarity of context between documents, in addition to the D2Q similarity.}
	\label{fig:D2D_example}
\end{figure}

\section{The D2D-based Method} \label{sec:approach}
In this section, we first attempt to explain the reason why the D2D-based approach can deal with the ``multiple degrees of similarity'' problem compared to D2Q-based approach. Then we introduce two methods to generate embeddings for documents and how to utilize embedding-based D2D similarities for document ranking.

\subsection{D2D Vs. D2Q Similarities} \label{subsec:D2D_example}
In this section, we attempt to explain the reason why the proposed D2D-based approach can improve over the D2Q similarity approach through example as in Figure \ref{fig:D2D_example}. Given a query `` Summer Olympics'' denoted as $Q$, there are two documents $d_1$ and $d_2$. $d_1$ is right about ``Summer Olympics'' while $d_2$ is about ``Winter Olympics''. Each document is assumed to consist of two parts, the keywords that summarize the content of the document, and the rest of the content, namely the context\footnote{We assume that the rest of the content in a document surrounds the keywords, to which the distance in word tokens could be long.}. Since ``Summer Olympics'' and ``Winter Olympics'' are very close in the embeddings space, both $d_1$ and $d_2$ are likely to be highly ranked to $Q$ using the D2Q approach as in \cite{Vuli2015Monolingual} which is based on the semantic similarities between queries and documents. However, it is obvious that $d_2$ is about ``Winter Olympics'' which is not relevant to Q. As the embeddings generated by Word or Para2Vec can be added together \cite{Mikolov2013a}, the embeddings of a document can also be seen as the summation of the embeddings of the keywords and the context. 

Let $KWs$ be the keywords and $Ctx$ be the context of $KWs$. Thus the embedding representation of a given document is as follows:
\begin{eqnarray}
	\label{eq:d}
	\vec{d_i}&=\vec{KWs_i} + \vec{Ctx_i} 
\end{eqnarray}

Let $\vec{Q}$ be the embedding of query $Q$. According to \cite{Vuli2015Monolingual}, the semantic relevance score based on the D2Q similarity is:
\begin{eqnarray}
	\label{eq:qd_score}
	SEM_{d\text{-}Q}(d_i, Q)=Cos(\vec{Q}, \vec{KWs_i}) + Cos(\vec{Q}, \vec{Ctx_i})
\end{eqnarray}
\noindent where the context within a document is matched to the query. Note that all embeddings are normalized to unit vectors since cosine similarity is used.

Instead of considering the similarity between a query and a document, semantic relevance score in our proposed approach is based on the similarities between a given document and highly relevant documents, simulated by the pseudo relevance set. According to our proposed approach, for a given relevant document $d_R$, the semantic relevance score of $d_i$ is:
\begin{eqnarray}
	\label{eq:dd_score}
	\begin{split}
		SEM_{d\text{-}d}(d_i, Q)&=\vec{d_i}\cdot\vec{d_R} \\
		&=(\vec{KWs_i} + \vec{Ctx_i})\cdot(\vec{KWs_R} + \vec{Ctx_R}) \\
		&=\vec{KWs_R}\cdot(\vec{KWs_i}+\vec{Ctx_i}) + \vec{Ctx_R}\cdot(\vec{KWs_i}+\vec{Ctx_i})
	\end{split}
\end{eqnarray}

Since $d_R$ is relevant to $Q$, $KWs_R$ is very close to $Q$ in the embeddings space. Therefore, Equation (\ref{eq:dd_score}) can be approximated as follows:
\begin{eqnarray}
	\label{eq:dd1_score}
	\begin{split}
		SEM_{d\text{-}d}(d_i, Q)& \approx \vec{Q}\cdot(\vec{KWs_i}+\vec{Ctx_i}) + \vec{Ctx_R}\cdot(\vec{KWs_i}+\vec{Ctx_i}) \\
		&=SEM_{d\text{-}Q}(d_i, Q) + \vec{Ctx_R}\cdot\vec{KWs_i} + \vec{Ctx_R}\cdot\vec{Ctx_i}
	\end{split}
\end{eqnarray}
%Be similar to Equation (\ref{eq:dd1_score}), the semantic relevance score of $d_2$ for given relevant document $d_R$ is:
%\begin{eqnarray}
%\label{eq:dd2_score}
%SEM_{d-d}(d_2, Q)=SEM_{d-Q}(d_2, Q) + \vec{Ctx_R}\cdot\vec{Kws_2} + \vec{Ctx_R}\cdot\vec{Ctx_2}
%\end{eqnarray}
Considering that $KWs_1$ and $KWs_2$ are very close to each other in the embeddings space, we have $\vec{Ctx_R} \cdot \vec{KWs_1} \approx \vec{Ctx_R} \cdot \vec{KWs_2}$. The above formula can be simplified by removing $\vec{Ctx_R} \cdot \vec{KWs_i}$ as follows:. %Considering that $\vec{KWs_i}$ and $\vec{KWs_j}$ are both close to $\vec{Q}$, we can simplify $SEM_{d-d}(d_1, Q)$ and $SEM_{d-d}(d_2, Q)$ as follows without changing the relative size between them:
\begin{eqnarray}
	\label{eq:dd12_score}
	\begin{split}
		SEM_{d\text{-}d}(d_1, Q)=SEM_{d\text{-}Q}(d_1, Q) + Cos(\vec{Ctx_R}, \vec{Ctx_1}) \\
		SEM_{d\text{-}d}(d_2, Q)=SEM_{d\text{-}Q}(d_2, Q) + Cos(\vec{Ctx_R}, \vec{Ctx_2})
	\end{split}
\end{eqnarray}

According to Equation (\ref{eq:dd12_score}), the D2D similarity considers the similarity of context between documents, in addition to the D2Q similarity. This is beneficial because queries are usually much shorter than documents, and the summation of query term embeddings does not necessarily cover sufficient semantic information of relevant documents. On the other hand, the use of D2D similarity enriches the semantic information involved in the document ranking. Continue with the example above, because both $Ctx_1$ and $Ctx_R$ are about sports of ``Summer Olympics'' while $Ctx_2$ is about sports of ``Winter Olympics'', we can tell that $d_1$ is likely to receive a higher similarity with $d_R$ than $d_2$. For this reason, the proposed approach in this paper has the potential to deal with the ``multiple degrees of similarity'' problem.

%According to Equation (\ref{eq:dd12_score}), our proposed approach also considers the similarities between the contexts of the keywords for a given document and a relevant document compared to \cite{Vuli2015Monolingual}. Because of the fact that both $Ctx_1$ and $Ctx_R$ are about sports of ``Summer Olympics'' while $Ctx_2$ is about the sports of ``Winter Olympics'', we can tell that $d_1$ is much more relevant than $d_2$. For this reason, the proposed approach in this paper has the ability to deal with the ``multiple degrees of similarity'' problem. 

\subsection{Generating Document Embeddings}
\label{sec:embeddings}

%The recently proposed neural embedding framework Word or Para2Vec \cite{Mikolov2013a,DBLP:journals/corr/LeM14} has been shown to be effective and efficient in many NLP tasks. In addition, Word or Para2Vec embedding technique has also been applied in information retrieval
% with positive results\cite{Ai2016Improving,Navid2016Generalizing,Vuli2015Monolingual}. Thus we are going to utilize the Word or Para2Vec embedding framework proposed by Mikolov et al. \cite{Mikolov2013a,DBLP:journals/corr/LeM14} to generate embeddings of words and documents.

A unique advantage of Word or Para2Vec is that the word embeddings are able to preserve semantic relationships on vector operations such as addition and subtraction \cite{Mikolov2013a}. For this reason, the first method we adopt to generate embeddings for documents is to sum up the embeddings of all the terms in a given document in a weighted manner. This method is denoted as \textit{Term Addition}, whose formula is given as follows: 

%A major challenge of the application of the word embeddings to IR is how to generate the document embeddings that are effective enough for document ranking. In this paper, we adopt two ways of generating document embeddings, namely \textit{Term Addition} and \textit{Paragraph Vector}. As the semantic relationships are preserved on the vector operations, one way of generating document embeddings is to add the word embeddings of the top-k most informative words in documents, i.e. \textit{Term Addition}, which is given by:
\begin{eqnarray}
	\label{eq:doc_embedding_add}
	\vec{d}=\sum_{w \in D_{t}} tf\text{-}idf(w) \cdot \vec{w}
\end{eqnarray}
\noindent where $\vec{w}$ and $\vec{d}$ are the embeddings of word $w$ and document $d$, respectively. $D_{t}$ is the set of all the terms in document $d$. $tf\text{-}idf(w)$ is used to measure the amount of information carried by word $w$, which is given by:
\begin{eqnarray}
	tf\text{-}idf(w)=tf \cdot \log_2 \frac{N-df_w+0.5}{df_w+0.5}
\end{eqnarray}
\noindent where $tf$ is the term frequency of $w$ in document $d$. $N$ is the total number of documents in the collection, and $df_w$ is the document frequency of word $w$.

Since the embeddings themselves learned by Paragraph Vector techique are representations of documents, we just directly utilize Paragraph Vector technique as the second method to generate embeddings for documents, which is denoted as \textit{Paragraph Vector}. 
In order to distinguish between the document embeddings generated by the two above mentioned methods, we denote document embeddings generated by \textit{Term Addition} as $\vec{d}_{add}$, and embeddings generated by \textit{Paragraph Vector} as $\vec{d}_{pv}$, where $add$ stands for \textit{Term Addition} and $pv$ stands for \textit{Paragraph Vector}.

%In Section \ref{sec:exp_settings}, the above two methods are used to generate embeddings for documents and queries. 

\subsection{Using D2D similarities for Document Ranking}
\label{sec:framework}
%As described in Section \ref{sec:introduction}, the use of the semantic similarity of the query-document pair may lead to limited improvement in retrieval performance, due to the inability in overcoming the problem of ``multiple degrees of similarity''. To deal with this problem, we integrate D2D-based semantic relevance score by measuring the semantic similarity between a document and the corresponding pseudo relevance feedback set. 

The ranking function of our proposed D2D approach is as follows:
\begin{eqnarray}
	\label{eq:aver_sim}
	score(d,Q)&=&\lambda R(d,Q)\\ \nonumber
	& &+(1-\lambda)SEM(d,D^k_{PRF}(Q))
\end{eqnarray}
\noindent where $R(d,Q)$ is the relevance score of document $d$ for a query Q given by a baseline model such as QLM with RM3. $D^k_{PRF}(Q)$ is the pseudo feedback set, which consists of the top-k documents according to $R(d,Q)$\footnote{Note that the pseudo feedback set here refers to the top-k documents used for the estimation of the $SEM(d,D^k_{PRF}(Q))$ score, which is different from the PRF method used in the baseline model, such as RM3.}. It is usually assumed by the PRF technique that most of the documents in $D^k_{PRF}(Q)$ are relevant to $Q$. Thus $D^k_{PRF}(Q)$ can be utilized to simulate the highly relevant document set. $SEM(d,D^k_{PRF}(Q))$ is the semantic relevance score of document $d$ for given pseudo relevance feedback set $D^k_{PRF}(Q)$. It is computed by the following formula: %\footnote{To avoid negative similarity, we change the formula slightly in experiments as follows:  $s=(D^TD+I)w$, in which $I$ is the identity matrix.}:
\begin{eqnarray}
	\label{eq:prf_sim}
	%SEM(d,D^k_{PRF}(Q))=\sum_{d' \in D^k_{PRF}} w_{d'} \cdot Sim(d',d)
	SEM(d,D^k_{PRF}(Q))={\bold{w}}^T({\bold{D}}^T\bold{d} + \bold{1})
\end{eqnarray}
\noindent where $\bold{D}$ is a $m \times k$ matrix, the i-th column of which is the embedding of the i-th document $d_i$ in $D^k_{PRF}(Q)$, normalized by its 2-norm. $\bold{w}$ is a $k \times 1$ vector, the i-th element of which is the weight corresponding to the document $d_i$ in $D^k_{PRF}(Q)$. $\bold{d}$ is a $m \times 1$ vector, namely the embedding of document $d$, also normalized by 2-norm. $\bold{1}$ is a $k \times 1$ vector whose elements are all 1, introduced to avoid negative similarity. Note that Min-Max normalization is applied on both of them to make $R(d, Q)$ and $SEM(d, D_{PRF}^k)$ be of the same scale.

The basic idea of Equation (\ref{eq:prf_sim}) is that the semantic similarity score of a given document $d$ is a linear combination of similarity scores between $d$ and the top-k feedback documents. The linear combination coefficient $\bold{w}$ reflects the weights of the feedback documents, which is given by the relevance score $R(d,Q)$.

\begin{table}[tbh]
	\caption{Statistics about the test collections and topics.}
	\label{tab:testColl}
	\centering
	\begin{tabular}{|c|p{2.0cm}<\centering|p{1.0cm}<\centering|p{1.0cm}<\centering|p{1.2cm}<\centering|}
		\hline
		Collection   		& TREC Task			&Topics	&\tabincell{c}{$\#$ of \\Topics}	&\tabincell{c}{$\#$ of \\Docs} \\
		\hline
		disk1\&2		&1, 2, 3 ad-hoc		&51-200& 150 	&741,856 \\ \hline
		disk4\&5		&Robust 2004		&\tabincell{c}{301-450\\601-700}& 250 &528,155\\ \hline
		WT10G		&9, 10 Web			&451-550& 100 &1,692,096 \\
		\hline
		GOV2		&\tabincell{c}{2004-2006 \\Terabyte Ad-hoc}	&701-850& 150&25,178,548 \\
		\hline
		ClueWeb09 B & 2009\text{-}2011 Web & wt1\text{-}150 & 150 & 50,220,423 \\ \hline
	\end{tabular}
\end{table}

%------------------------------------------------

\section{Experimental Settings} \label{sec:exp_settings}
In this section, we introduce our experimental settings, including datasets and the experimental design.

\subsection{Datasets}
\label{sec:datasets}
We use five standard TREC test collections in our experiments, and the basic statistics about the test collections and topics are given in Table \ref{tab:testColl}. The test collections are publicly available and have been widely used in evaluation of related approaches \cite{Ai2016Improving,DBLP:conf/sigir/GangulyRMJ15,DBLP:conf/cikm/GuoFAC16,DBLP:conf/cikm/KuziSK16,DBLP:journals/corr/RoyPMG16,DBLP:conf/ictir/ZamaniC16a,Zamani2016Estimating,Zheng2015Learning,DBLP:conf/cikm/GuoFAC16a}.

All experiments are conducted using a computer cluster with 4 nodes. Each node is equipped with 8G of RAM and four Intel i5-2400 cores running at 3.10 GHz. In addition, we use the open source Terrier toolkit version 4.2 \cite{Mccreadie2014Terrier} to index the test collections with the recommended settings of the toolkit. On the five test collections, documents are preprocessed by removing all HTML tags, standard English stopwords are removed and the test collections are stemmed using Porter's English stemmer. Each topic contains three fields, i.e. title, description and narrative, and we only use the title field. The title-only queries are very short which is usually regarded as a realistic snapshot of real user queries.

For each test collection, the Skip-gram model of Word or Para2Vec toolkit$^1$ with negative sampling \cite{DBLP:journals/corr/GoldbergL14} is utilized to generate word and document embeddings, which are trained by stochastic gradient ascent. Due to the exhaustive memory consumption during the training, the word or paragraph embeddings of each collection are learned from a subset of documents that is composed of the top-1000 documents returned by each test query. For example, for the 100 test queries on WT10G, 100K documents minus redundancies are merged to learn the embeddings. The window size is set to 10 for Skip-gram model as recommended by \cite{Mikolov2013a}. The number of dimensions of the embeddings are set to 300. In fact, according to Table \ref{tab:dim}, with a wide range of possible settings, changing the number of dimensions of the word and document embeddings has little impact on the retrieval performance.

% Table
\begin{table}[tbh]
	\centering
	\caption{Comparison to \textit{BM25}. The results of ClueWeb09B (CW09B) is reported on nDCG@20, and the rest are reported on MAP. A statistically significant difference is marked with a *. The best result on each collection is in \textbf{bold}.}
	\label{tab:bm25_noqe}
	%\begin{tabular}{|c||c|c|c|c|c|}
	\begin{tabular}{|p{1.5cm}<{\centering}||p{1.2cm}<{\centering}|p{1.2cm}<{\centering}|p{1.2cm}<{\centering}|p{1.2cm}<{\centering}|p{1.2cm}<{\centering}|}
		\hline
		Model & disk1\&2 & disk4\&5 & WT10G & GOV2 & CW09B \\ \hline \hline
		$\scriptstyle BM25$ & 0.2408 & 0.2534 & 0.2123 & 0.3008 & 0.2251 \\ \hline \hline
		$\scriptstyle BM25+SEM_{d_{LDA}}$ & \tabincell{c}{0.2517\\+4.53\%*} & \tabincell{c}{0.2675\\+5.56\%*} & \tabincell{c}{0.2158\\+1.65\%} & \tabincell{c}{0.3193\\+6.15\%*} & \tabincell{c}{0.2306\\+2.44\%*} \\ \hline
		$\scriptstyle BM25+SEM_{d_{TFIDF}}$ & \tabincell{c}{0.2477\\+2.87\%} & \tabincell{c}{0.2554\\+0.79\%} & \tabincell{c}{0.2187\\+3.01\%} & \tabincell{c}{0.3043\\+1.16\%} & \tabincell{c}{0.2311\\+2.67\%*} \\ \hline \hline
		$\scriptstyle BM25+SEM_{d_{pv}}$ & \tabincell{c}{\textbf{0.2820}\\\textbf{+17.11\%*}} & \tabincell{c}{\textbf{0.2862}\\\textbf{+12.94\%*}} & \tabincell{c}{\textbf{0.2427}\\\textbf{+14.32\%*}} & \tabincell{c}{0.3138\\+4.32\%*} & \tabincell{c}{\textbf{0.2452}\\\textbf{+8.93\%*}} \\ \hline
		$\scriptstyle BM25+SEM_{d_{add}}$ & \tabincell{c}{0.2727\\+13.25\%*} & \tabincell{c}{0.2796\\+10.34\%*} & \tabincell{c}{0.2423\\+14.13\%*} & \tabincell{c}{\textbf{0.3184}\\\textbf{+5.85\%*}} & \tabincell{c}{0.2404\\+6.80\%*} \\ \hline
		
	\end{tabular}
\end{table}

\begin{table}[tbh]
	\centering
	\caption{Comparison to \textit{BM25 with Rocchio's PRF ($\scriptstyle BM25_{PRF}$)}. The results of ClueWeb09B (CW09B) is reported on nDCG@20, and the rest are reported on MAP. A statistically significant difference is marked with a *. The best result on each collection is in \textbf{bold}.}
	\label{tab:bm25_prf}
	%\begin{tabular}{|c||c|c|c|c|c|}
	\begin{tabular}{|p{1.5cm}<{\centering}||p{1.2cm}<{\centering}|p{1.2cm}<{\centering}|p{1.2cm}<{\centering}|p{1.2cm}<{\centering}|p{1.2cm}<{\centering}|}
		\hline
		Model & disk1\&2 & disk4\&5 & WT10G & GOV2 & CW09B \\ \hline \hline
		$\scriptstyle BM25_{PRF}$ & 0.3083 & 0.2966 & 0.2445 & 0.3430 & 0.2536 \\ \hline \hline
		$\scriptstyle BM25_{PRF}+SEM_{d_{LDA}}$ & \tabincell{c}{0.3084\\+0.03\%} & \tabincell{c}{0.2966\\+0.0\%} & \tabincell{c}{0.2446\\+0.04\%} & \tabincell{c}{0.3479\\+1.43\%} & \tabincell{c}{0.2596\\+2.37\%*} \\ \hline
		$\scriptstyle BM25_{PRF}+SEM_{d_{TFIDF}}$ & \tabincell{c}{0.3093\\+0.32\%} & \tabincell{c}{0.2967\\+0.03\%} & \tabincell{c}{0.2445\\+0.0\%} & \tabincell{c}{0.3430\\+0.0\%} & \tabincell{c}{0.2587\\+2.01\%} \\ \hline \hline
		$\scriptstyle BM25_{PRF}+SEM_{d_{pv}}$ & \tabincell{c}{\textbf{0.3110}\\\textbf{+0.88\%}} & \tabincell{c}{\textbf{0.2990}\\\textbf{+0.81\%}} & \tabincell{c}{\textbf{0.2541}\\\textbf{+3.93\%*}} & \tabincell{c}{0.3484\\+1.57\%} & \tabincell{c}{0.2635\\+4.87\%*} \\ \hline
		$\scriptstyle BM25_{PRF}+SEM_{d_{add}}$ & \tabincell{c}{0.3105\\+0.71\%} & \tabincell{c}{0.2985\\+0.64\%} & \tabincell{c}{\textbf{0.2541}\\\textbf{+3.93\%*}} & \tabincell{c}{\textbf{0.3523}\\\textbf{+2.71\%*}} & \tabincell{c}{\textbf{0.2677}\\\textbf{+5.42\%*}} \\ \hline
		
	\end{tabular}
\end{table}

\begin{table}[tbh]
	\centering
	\caption{Comparison to \textit{QLM with RM3 ($\scriptstyle QLM_{PRF}$)}. The results of ClueWeb09B (CW09B) is reported on nDCG@20, and the rest are reported on MAP. A statistically significant difference is marked with a *. The best result on each collection is in \textbf{bold}.}
	\label{tab:qlm_prf}
	%\begin{tabular}{|l||c|c|c|c|c|}
	\begin{tabular}{|p{1.5cm}<{\centering}||p{1.2cm}<{\centering}|p{1.2cm}<{\centering}|p{1.2cm}<{\centering}|p{1.2cm}<{\centering}|p{1.2cm}<{\centering}|}
		\hline
		Model & disk1\&2 & disk4\&5 & WT10G & GOV2 & CW09B \\ \hline \hline
		$\scriptstyle QLM_{PRF}$ & 0.2691 & 0.2837 & 0.2369 & 0.3319 & 0.2278 \\ \hline \hline
		$\scriptstyle QLM_{PRF}+SEM_{d_{LDA}}$ & \tabincell{c}{0.2695\\+0.15\%} & \tabincell{c}{0.2902\\+2.29\%*} & \tabincell{c}{0.2370\\+0.04\%} & \tabincell{c}{0.3353\\+1.02\%} & \tabincell{c}{0.2311\\+1.45\%} \\ \hline
		$\scriptstyle QLM_{PRF}+SEM_{d_{TFIDF}}$ & \tabincell{c}{0.2693\\+0.07\%} & \tabincell{c}{0.2842\\+0.18\%} & \tabincell{c}{0.2369\\+0.0\%} & \tabincell{c}{0.3319\\+0.0\%} & \tabincell{c}{0.2315\\+1.62\%} \\ \hline \hline
		$\scriptstyle QLM_{PRF}+SEM_{d_{pv}}$ & \tabincell{c}{\textbf{0.2893}\\\textbf{+7.51\%*}} & \tabincell{c}{\textbf{0.2956}\\\textbf{+4.19\%*}} & \tabincell{c}{\textbf{0.2470}\\\textbf{+4.26\%*}} & \tabincell{c}{0.3359\\+1.21\%} & \tabincell{c}{\textbf{0.2423}\\+\textbf{6.37\%*}} \\ \hline
		$\scriptstyle QLM_{PRF}+SEM_{d_{add}}$ & \tabincell{c}{0.2850\\+5.91\%*} & \tabincell{c}{0.2929\\+3.24\%*} & \tabincell{c}{0.2445\\+3.21\%*} & \tabincell{c}{\textbf{0.3378}\\\textbf{+1.78\%*}} & \tabincell{c}{0.2401\\+5.40\%*} \\ \hline
		
	\end{tabular}
\end{table}

\subsection{Experimental Design}\label{sec:exp_design}
In our experiments, we evaluate our approach against the following strong baselines, i.e. BM25 \cite{robertson1996okapi}, BM25 with Rocchio's PRF method \cite{DBLP:conf/ictir/HuiHLW11}, QLM with RM3 \cite{lavrenko2001relevance}, and the D2Q-based approach proposed in \cite{Vuli2015Monolingual}. Note that the Rocchio's PRF method used in this paper is not the same as the original version as proposed in \cite{rocchio1971relevance}. It has been adopted for recent IR models, as in \cite{DBLP:conf/ictir/HuiHLW11}. BM25 with PRF is 
denoted as $BM25_{PRF}$, while QLM with PRF is denoted as $QLM_{PRF}$. In addition to the above baselines, the topic model LDA \cite{blei2003latent}, and TF-IDF \cite{Manning2008ir} are compared to Word or Para2Vec in generating the vector representations of documents in our experiments.

The baseline models used in our experiments are optimized by grid search algorithm \cite{Bergstra2011Algorithms}. As described in Section \ref{sec:embeddings}, we use two methods to generate document embeddings denoted as ${\vec{d}_{add}}$ and ${\vec{d}_{pv}}$, respectively. The semantic relevance score generated by our approach is denoted as $SEM_{d_{pv}}$ or $SEM_{d_{add}}$ depending on the embeddings used, while $Sim_{d\text{-}Q}$ denotes the semantic relevance score generated by the D2Q-based approach in \cite{Vuli2015Monolingual}. For example, if we use BM25 to generate the content-based relevance score, and use \textit{Paragraph Vector} to generate embeddings for documents, our approach is denoted as $BM25+SEM_{d_{pv}}$. Our approach has the following tunable parameters, hyper-parameter $\lambda$ (see Equation (\ref{eq:aver_sim})), and top $k$ documents in $D^k_{PRF}(Q)$ ($|D^k_{PRF}|$). %Besides, in Section \ref{sec:comparison}, our approach is compared to some recently proposed approaches.

On each collection, we evaluate by a two-fold cross-validation. The queries for each test collection are split into two equal-size subsets by parity in odd or even topic numbers. 
In each fold, one subset is used for training, and the other is used for test. The results reported in the paper are averaged over queries in the two test subsets. There is no overlap between the training and test subsets. We report on the official TREC evaluation metrics, including Mean Average Precision (\textit{MAP}) \cite{Gonzalez2007jasist} on disk1\&2, disk4\&5, WT10G, and GOV2, and \textit{nDCG@20} \cite{Gonzalez2007jasist} on ClueWeb09 B. We use the official TREC evaluation metrics as we trust the TREC organizers to pick the appropriate measures for different retrieval tasks. In addition, in a comparison to the state-of-the-art method in \cite{DBLP:conf/acl/0001MC16}, \textit{nDCG@10} is used instead, since it is the reported metric in \cite{DBLP:conf/acl/0001MC16}. All statistical tests are based on the t-test at the 0.05 significance level, which is a popular setting used in a number of recent related studies \cite{DBLP:conf/cikm/RoyGMJ16}\cite{DBLP:journals/corr/RoyPMG16}\cite{DBLP:conf/ictir/ZamaniC16a}.

\begin{table}[tbh]
	\centering
	\caption{Comparison to \textit{$\scriptstyle BM25+Sim_{d\text{-}Q}$} \cite{Vuli2015Monolingual}. The results of ClueWeb09B (CW09B) is reported on nDCG@20, and the rest are reported on MAP. A statistically significant difference is marked with a *. The best result on each collection is in \textbf{bold}.}
	\label{tab:bm25_noqe_qd}
	%\begin{tabular}{|c||c|c|c|c|c|}
	\begin{tabular}{|p{1.0cm}<{\centering}||p{1.2cm}<{\centering}|p{1.2cm}<{\centering}|p{1.2cm}<{\centering}|p{1.2cm}<{\centering}|p{1.2cm}<{\centering}|}
		\hline
		Model & disk1\&2 & disk4\&5 & WT10G & GOV2 & CW09B \\ \hline \hline
		$\scriptstyle BM25+Sim_{d\text{-}Q}$ & 0.2547 & 0.2598 & 0.2196 & 0.3072 & 0.2278 \\ \hline \hline
		$\scriptstyle BM25+SEM_{d_{pv}}$ & \tabincell{c}{\textbf{0.2820}\\\textbf{+10.72\%*}} & \tabincell{c}{\textbf{0.2862}\\\textbf{+10.16\%*}} & \tabincell{c}{\textbf{0.2427}\\\textbf{+10.52\%*}} & \tabincell{c}{0.3138\\+2.15\%} & \tabincell{c}{\textbf{0.2452}\\\textbf{+7.64\%*}} \\ \hline
		$\scriptstyle BM25+SEM_{d_{add}}$ & \tabincell{c}{0.2727\\+7.07\%*} & \tabincell{c}{0.2796\\+7.62\%*} & \tabincell{c}{0.2423\\+10.34\%*} & \tabincell{c}{\textbf{0.3184}\\\textbf{+3.65\%*}} & \tabincell{c}{0.2404\\+5.53\%} \\ \hline
		
	\end{tabular}
\end{table}

\begin{table}[tbh]
	\centering
	\caption{Comparison to \textit{$\scriptstyle QLM_{PRF}+Sim_{d\text{-}Q}$} \cite{Vuli2015Monolingual}. The results of ClueWeb09B (CW09B) is reported on nDCG@20, and the rest are reported on MAP. A statistically significant difference is marked with a *. The best result on each collection is in \textbf{bold}.}
	\label{tab:qlm_prf_qd}
	%\begin{tabular}{|c||c|c|c|c|c|}
	\begin{tabular}{|p{1.5cm}<{\centering}||p{1.2cm}<{\centering}|p{1.2cm}<{\centering}|p{1.2cm}<{\centering}|p{1.2cm}<{\centering}|p{1.2cm}<{\centering}|}
		\hline
		Model & disk1\&2 & disk4\&5 & WT10G & GOV2 & CW09B \\ \hline \hline
		$\scriptstyle QLM_{PRF}+Sim_{d\text{-}Q}$ & 0.2759 & 0.2859 & 0.2375 & 0.3319 & 0.2324 \\ \hline \hline
		$\scriptstyle QLM_{PRF}+SEM_{d_{pv}}$ & \tabincell{c}{\textbf{0.2893}\\\textbf{+4.86\%*}} & \tabincell{c}{\textbf{0.2956}\\\textbf{+3.39\%*}} & \tabincell{c}{\textbf{0.2470}\\\textbf{+4.00\%*}} & \tabincell{c}{0.3359\\+1.21\%} & \tabincell{c}{\textbf{0.2423}\\+\textbf{4.26\%*}} \\ \hline
		$\scriptstyle QLM_{PRF}+SEM_{d_{add}}$ & \tabincell{c}{0.2850\\+3.30\%*} & \tabincell{c}{0.2929\\+2.45\%*} & \tabincell{c}{0.2445\\+2.95\%} & \tabincell{c}{\textbf{0.3378}\\\textbf{+1.78\%*}} & \tabincell{c}{0.2401\\+3.31\%*} \\ \hline
		
		%text{-}D^k_{PRF}
		
	\end{tabular}
\end{table}

\section{Evaluation Results}
\label{sec:eval_results}
Table \ref{tab:bm25_noqe} presents the results against the classical BM25 model. According to the results, the integration of semantic relevance score (i.e. $SEM$) has statistically significant improvements over BM25 in all cases, indicating the effectiveness of our approach.

Tables \ref{tab:bm25_prf} and \ref{tab:qlm_prf} present the evaluation results against BM25 with Rocchio's PRF method and QLM with the RM3 relevance model, respectively. It is encouraging to see that statistically significant improvements are still observed with the use of PRF in most cases, especially on the three Web collections, showing the effectiveness of our approach.

Tables \ref{tab:bm25_noqe}\text{-}\ref{tab:qlm_prf} also present the comparison of three different models (i.e. Word or Para2Vec, LDA and TF-IDF) in generating the vector representations of documents. Out of the three models for document vector generating, Word or Para2Vec achieves the best effectiveness. LDA outperforms TF-IDF, but both of them are not as effective as Word or Para2Vec. The comparison result between Word or Para2Vec and LDA is consistent with the findings in other NLP tasks \cite{sun2016semantic,dai2015document}. As the TF-IDF vector representations of documents do not have the ability in capturing the semantic relation between texts, the comparison results between TF-IDF and the other two models can be expected.

The comparison between our approach and the approach proposed in \cite{Vuli2015Monolingual} is presented in Tables \ref{tab:bm25_noqe_qd} and \ref{tab:qlm_prf_qd}. According the results, our approach outperforms the D2Q-based approach proposed in \cite{Vuli2015Monolingual} in all cases, in which the semantic relevance of a document is measured by the cosine similarity between the embeddings of the document and the given query. As explained in Section \ref{subsec:D2D_example}, our proposed D2D-based approach has the ability to deal with the problem of ``multiple degrees of similarity'' when using embeddings to IR, and therefore the improvement over the D2Q approach in \cite{Vuli2015Monolingual} is expected.

\begin{figure*}[tbh]
	\footnotesize
	\centering
	\subfigure[Sensitivity to $\lambda$ when using $\vec{d}_{add}$]{\includegraphics[width=0.4\textwidth]
		{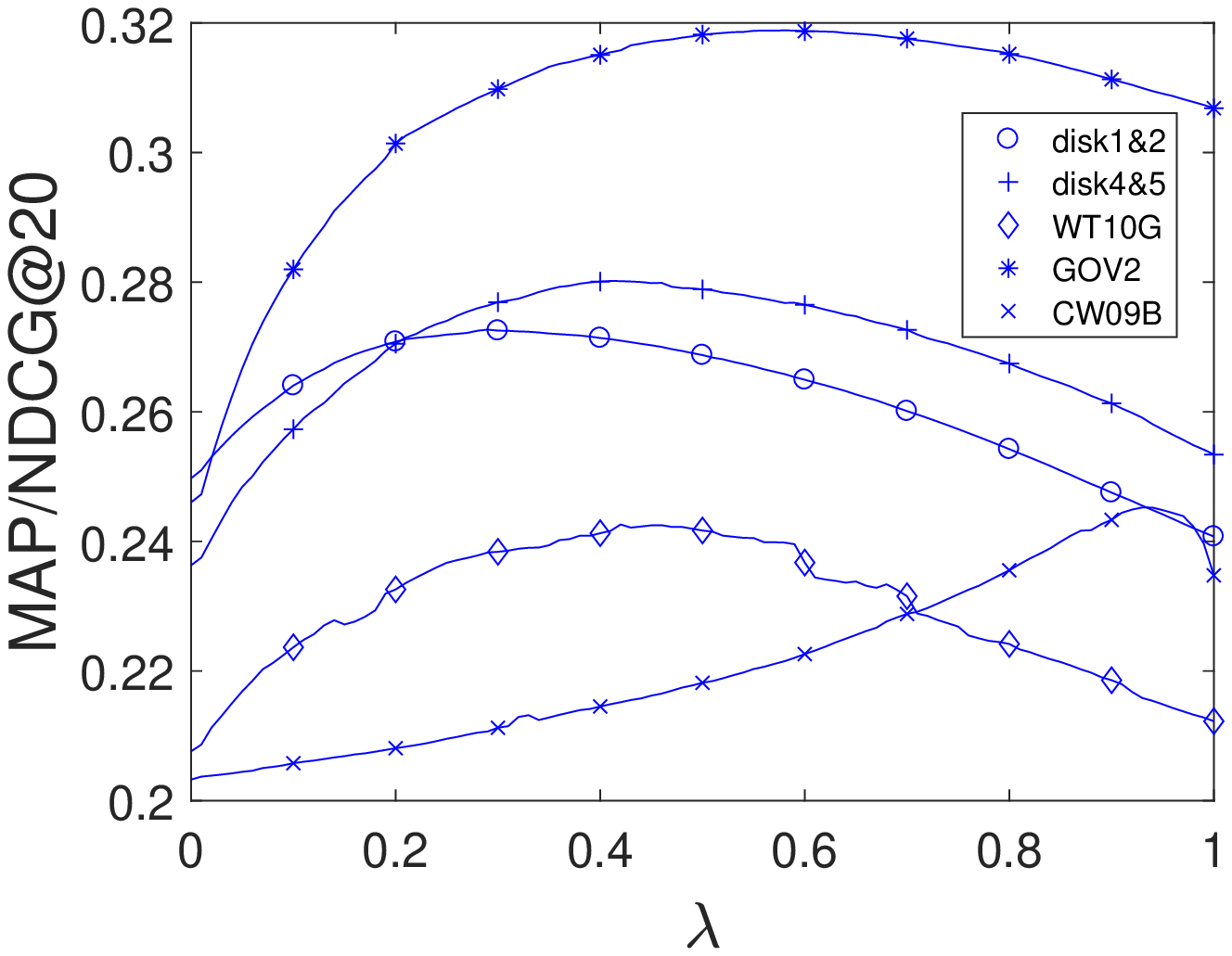} }
	\subfigure[Sensitivity to $\lambda$ when using $\vec{d}_{pv}$]{\includegraphics[width=0.4\textwidth]
		{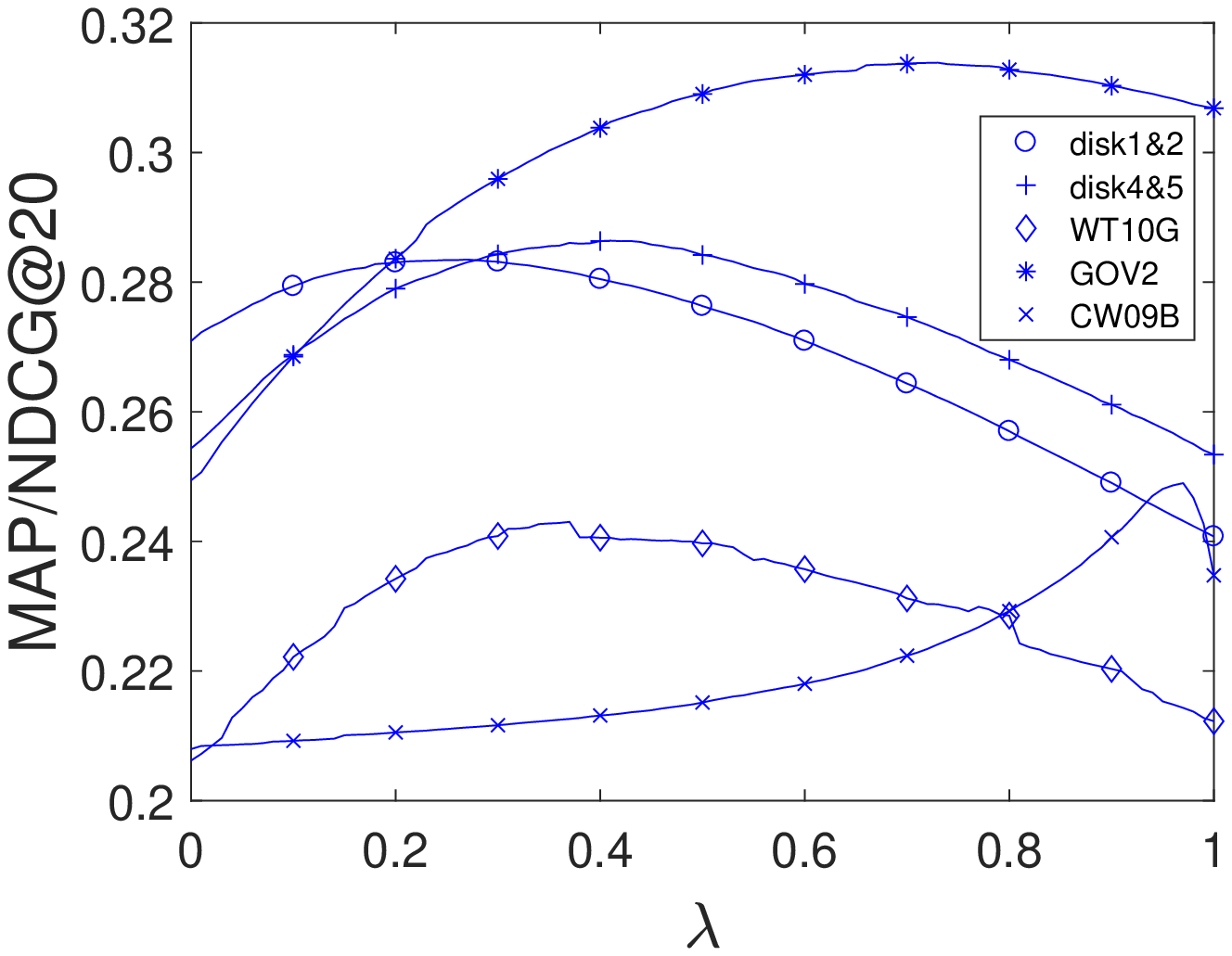} } %\\
	\subfigure[Sensitivity to $\scriptsize|D^k_{PRF}|$ when using $\vec{d}_{add}$]{\includegraphics[width=0.4\textwidth]
		{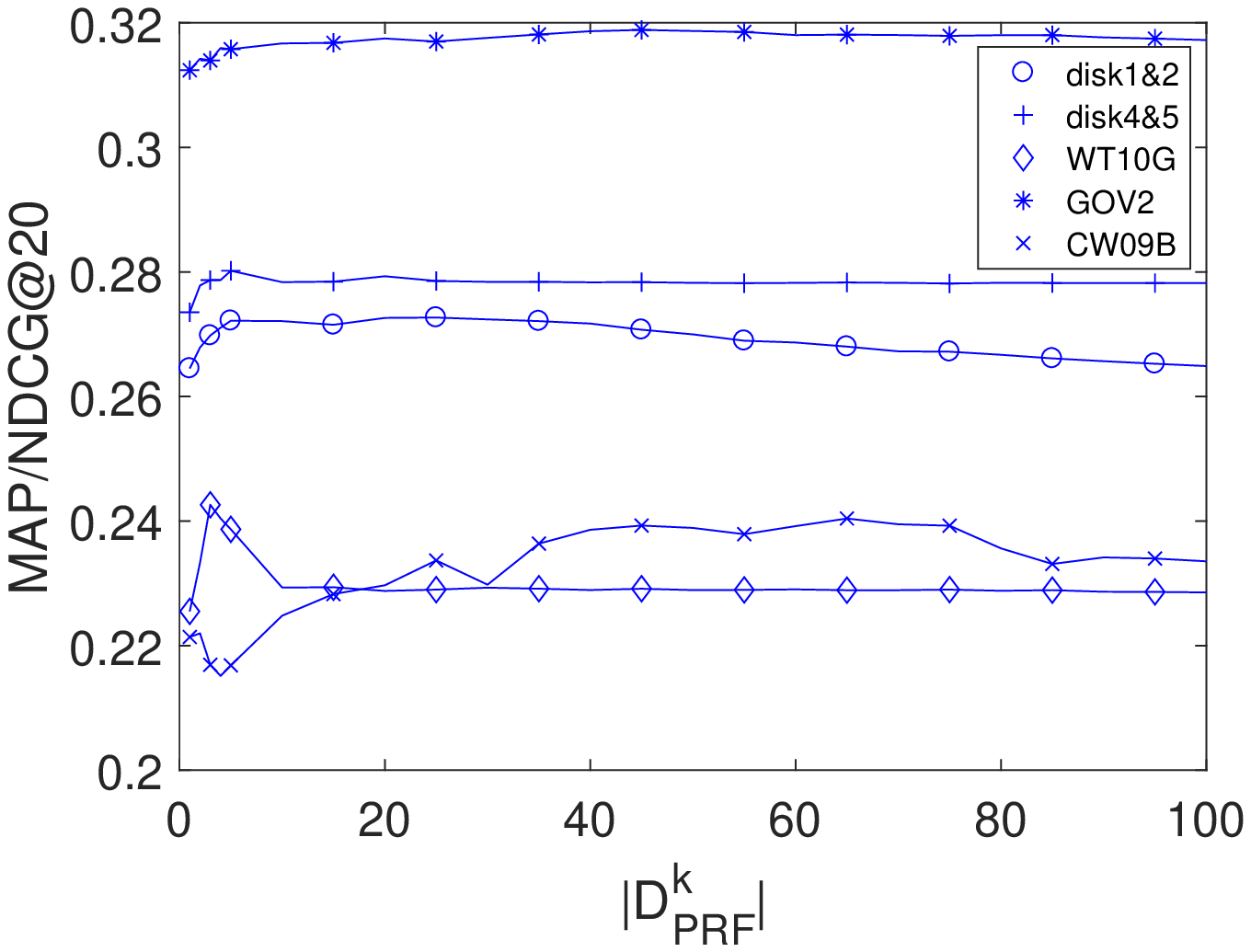} }
	\subfigure[Sensitivity to $\scriptsize|D^k_{PRF}|$ when using $\vec{d}_{pv}$]{\includegraphics[width=0.4\textwidth]
		{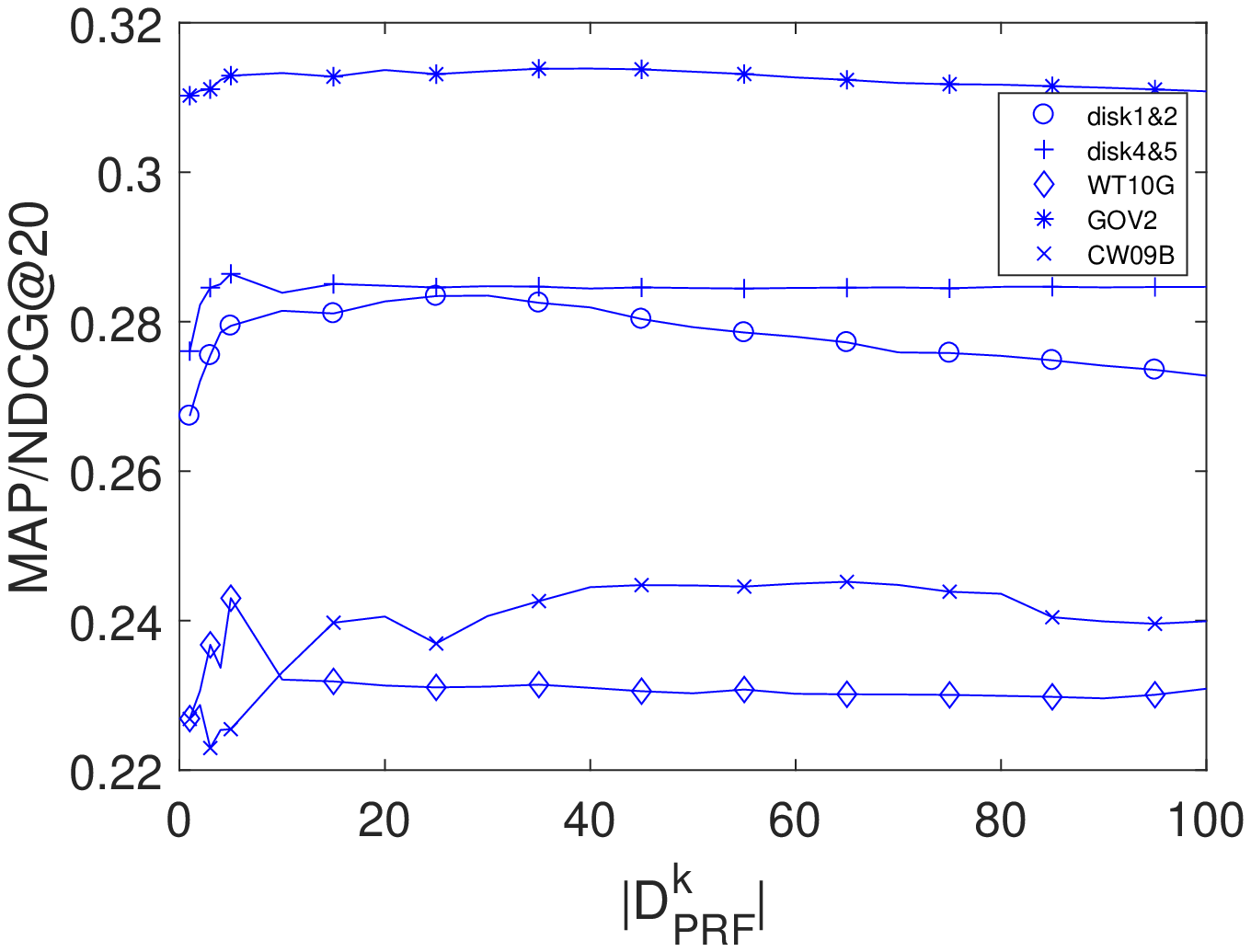} }
	
	\caption{The sensitivities of our approach to parameter $\lambda$ and $|D^k_{PRF}|$ when using $\vec{d}_{add}$ or $\vec{d}_{pv}$.}
	\label{fig:sens_para}
\end{figure*}

\begin{figure*}[tbh]
	\footnotesize
	\centering
	\subfigure[disk1\&2]{\includegraphics[width=0.4\textwidth]
		{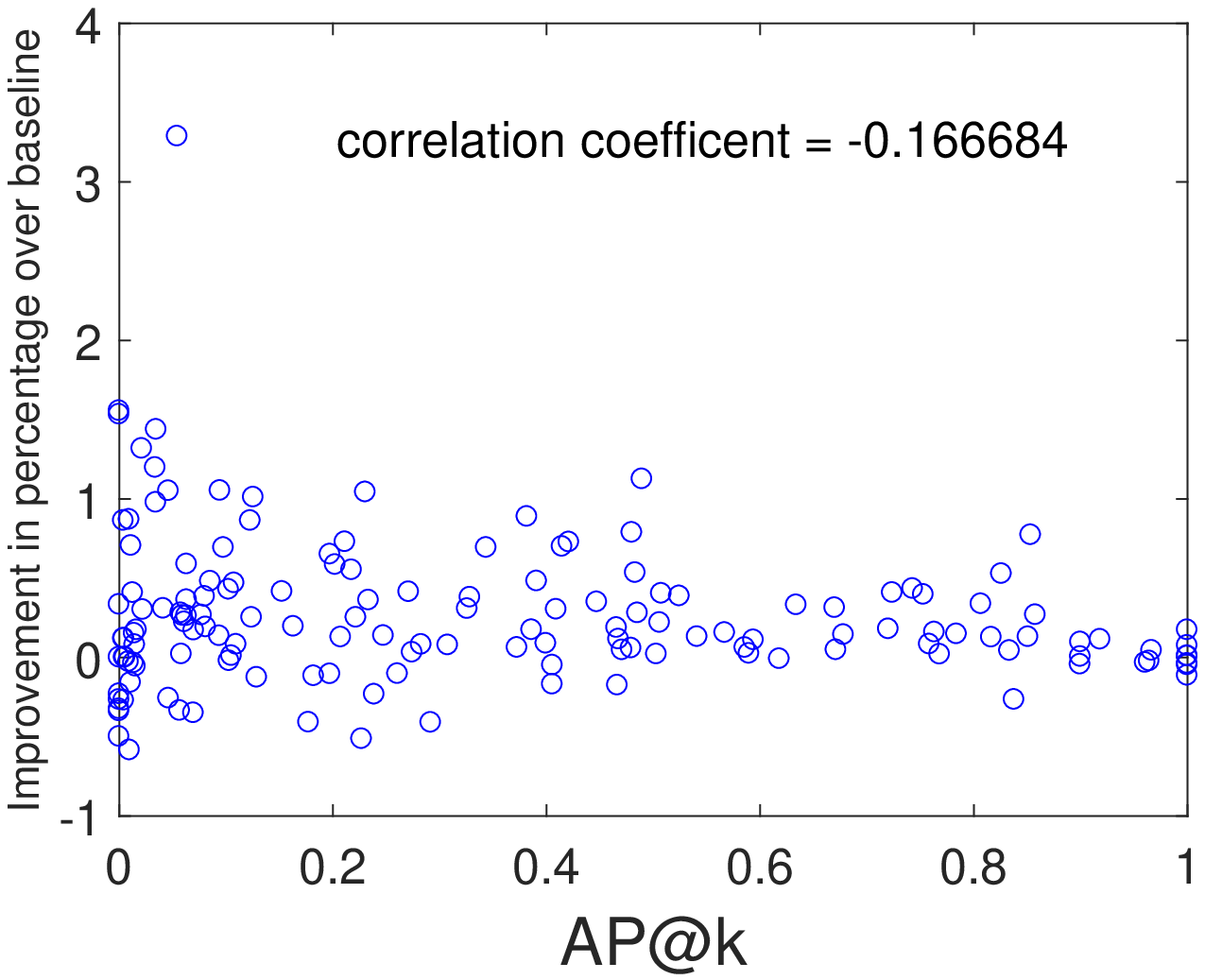} }
	\subfigure[disk4\&5]{\includegraphics[width=0.4\textwidth]
		{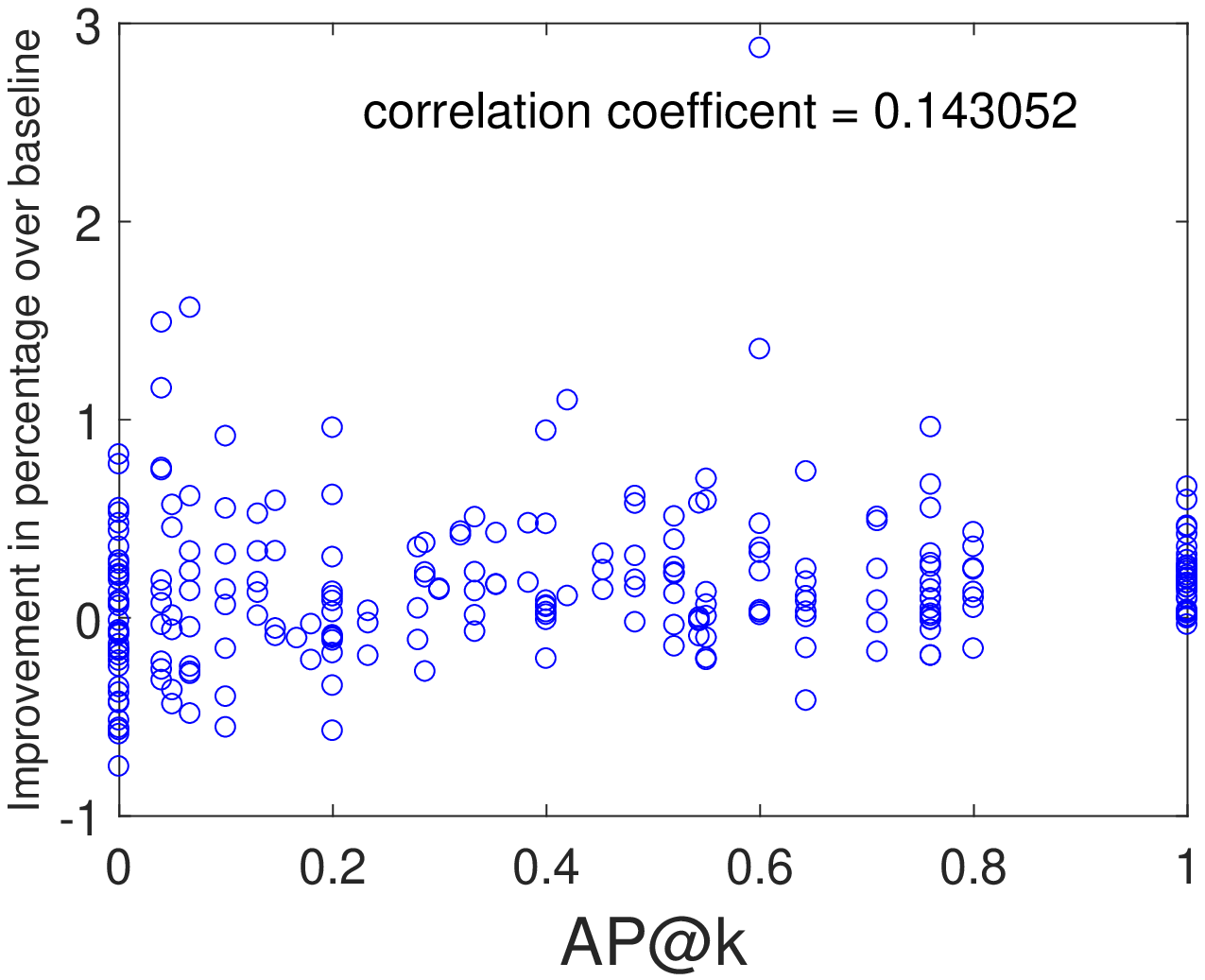} } %\\
	\subfigure[WT10G]{\includegraphics[width=0.4\textwidth]
		{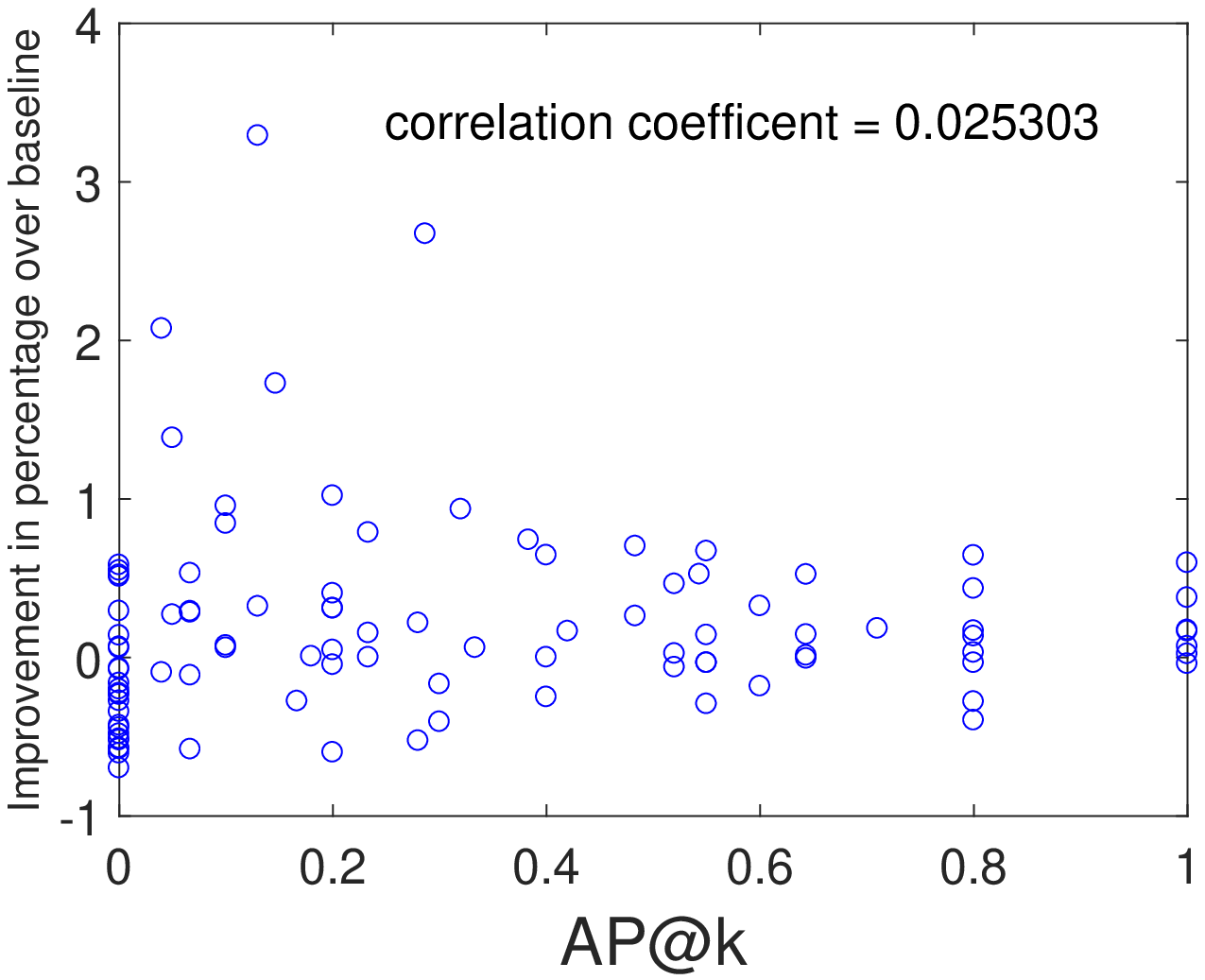} }
	\subfigure[GOV2]{\includegraphics[width=0.4\textwidth]
		{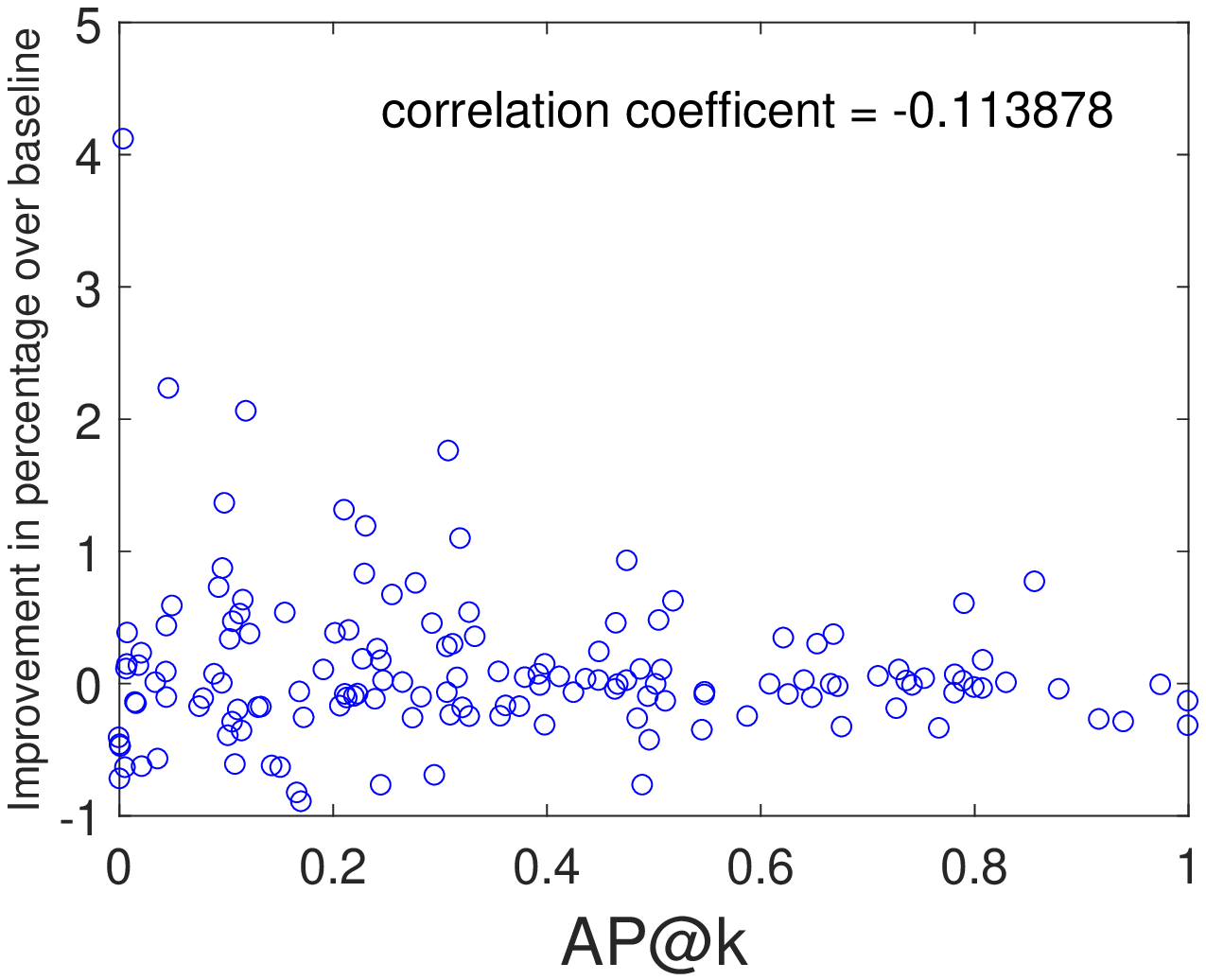} }
	
	\caption{The correlation between our approach's improvement in percentage over baseline and quality of the pseudo relevance set. The linear correlation obtained on ClueWeb09 B is 0.1536. Results on ClueWeb09 B are not plotted for brevity.}
	\label{fig:sens_prf}
\end{figure*}

%\begin{comment}
\begin{table*}[!tbh]
	\centering
	\caption{The impact of embedding dimensionality on retrieval performance obtained on WT10G by $BM25_{PRF}+SEM_{d_{add}}$. Parameter $b$ in BM25 is set to 0.60 \textbf{with PRF}.}
	\label{tab:dim}
	\begin{tabular}{|c|c|c|c|c|c|}
		\hline
		\#Dimensions & 100 & 200 & 300 & 400 & 500  \\ \hline
		MAP 		& 0.2371 & 0.2380 & 0.2382 & 0.2380 & 0.2389 \\ \hline\hline
		\#Dimensions & 600 & 700 & 800 & 900 & 1000 \\ \hline
		MAP 	& 0.2387 & 0.2391 & 0.2388 & 0.2393 & 0.2392 \\ \hline
		
	\end{tabular}
\end{table*}
%\end{comment}

\section{Further Evaluation and Analysis}\label{sec:discuss}
In this section, we study the sensitivity of our approach to the tunable parameters, the impact of the quality of pseudo relevance document set on our approach, compare our approach with recently proposed state-of-the-art approaches, and apply the proposed approach to the Clinical Decision Support (CDS) task.

%\begin{comment}
\subsection{Parameter Sensitivity Analysis}
As described in Section \ref{sec:exp_design}, our approach has the following tunable parameters, hyper-parameter $\lambda$ (see Equation (\ref{eq:aver_sim})) and top $k$ documents in $D^k_{PRF}(Q)$ ($|D^k_{PRF}|$). Embedding dimensionality (\textit{\# Dimensions}) is also a hyper-parameter in our approach. Note that the presented results in this section are all based on $BM25$. Similar results can be observed using $QLM$.  %We only present figures obtained by BM25, evaluated by MAP, as other experiments have similar observations.

For hyper-parameter $\lambda$, grid search algorithm is utilized to find the optimal value of $\lambda$ with an interval of 0.01 from 0 to 1. According to Equation (\ref{eq:aver_sim}), when $\lambda$ is set to 1, only $BM25$ is used. When $\lambda$ is set to 0, only the semantic relevance score (SEM) is used. (a) and (b) of  Figure \ref{fig:sens_para} present the sensitivity of our approach to hyper-parameter $\lambda$ on \textit{MAP} without th use of PRF when using $\vec{d}_{add}$ and $\vec{d}_{pv}$, respectively. From (a) and (b) of  Figure \ref{fig:sens_para} we can see that, the optimal values of $\lambda$ usually fall in the interval of $[0.25,0.45]$ in most cases. As hyper-parameter $\lambda$ controls the influence of the content-based model and the semantic relevance score shown in Equation (\ref{eq:aver_sim}), the optimal values of $\lambda$ in Figure \ref{fig:sens_para} indicate that the impact of semantic relevance score should be greater than $BM25$. In addition, we can find that the optimization curves of \textit{Paragraph Vector} and \textit{Term Addition} follow similar trends.

According to (c) and (d) of Figure \ref{fig:sens_para}, the optimal values of parameter $|D^k_{PRF}|$ are usually set to less than 50. As the parameter $|D^k_{PRF}|$ to a certain extent determines the overall semantic relevance represented by the pseudo feedback set ($D^k_{PRF}(Q)$), the parameter is quite critical to our approach. The steep optimization curves in (c) and (d) of Figure \ref{fig:sens_para} indicate that our approach is sensitive to parameter $|D^k_{PRF}|$, especially when the parameter is set to relatively small values. Once again, the optimization curves of \textit{Paragraph Vector} and \textit{Term Addition} follow similar trends in (c) and (d) of Figure \ref{fig:sens_para}. The results obtained with the use of PRF for the baseline have similar observations, which are not presented for brevity.

%There is an additional parameter that needs to tune, namely \textit{\# Terms}, when applying \textit{Term Addition} to generate embeddings for documents. For space reason, we do not present the results of the sensitivity study of \textit{\# Terms}. When not using PRF, the optimal values of \textit{\# Terms} often fall in the interval of $[30,80]$. An interesting observation is that the optimal values of \textit{\# Terms} are the number of terms in documents when PRF is used, indicating that the embedding of a document can be simply generated by summing up the embeddings of all the terms in the document.    % Figures \ref{fig:add_noqe} and \ref{fig:add_qe} present the sensitivity of parameters \textit{\# Terms} and \textit{\# PRF Documents} when applying \textit{Term Addition}. From Figures \ref{fig:add_noqe} and \ref{fig:add_qe} we can see that the sensitivity of parameter \textit{\# PRF Documents} when applying \textit{Paragraph Vector} or \textit{Term Addition} is almost similar. It is worth noting that all the term embeddings in a document are added to generate the document embedding when the parameter \textit{\# Terms} is set to 0. According to Figures \ref{fig:add_noqe} and \ref{fig:add_qe}, the optimal value of parameter \textit{\# Terms} is 0 in most cases, indicating that the embedding of a document can be simply generated by summing up the embeddings of all the terms in the document.

In addition, in order to analyze the impact of embedding dimensionality on the retrieval performance, we train word and document embeddings by setting a dimensionality interval of $[100,1000]$, with step 100. According to Table \ref{tab:dim}, with a wide range of possible settings, the number of dimensions of the word and document embeddings has little impact on the retrieval performance, showing that our approach has very stable retrieval effectiveness with respect to the change of the number of dimensions of the embeddings.
%\end{comment}

\begin{table}[!tbh]
	\caption{The mean of improvement in percentage over baseline at different levels of pseudo relevance set quality.}
	\label{tab:ap_k}
	\centering
	\begin{tabular}{|c|c|c|c|}
		\hline
		collection  &  $AP@k < 0.1$ & $AP@k \in [0.1,0.9]$	& $AP@k > 0.9$ \\ \hline
		disk1\&2 & 35.40\% & 24.00\% & 2.15\% \\ \hline
		disk4\&5 & 7.26\% & 18.48\% & 20.42\% \\ \hline
		WT10G	& 00.93\% & 30.70\% & 19.39\% \\ \hline
		GOV2 & 24.42\% & 7.43\% & -20.70\% \\ \hline
		CW09B & 10.45\% & 23.16\% &  8.44\% \\ \hline
	\end{tabular}
\end{table}

\subsection{Influence of the Quality of the Pseudo Feedback Set}
\label{sec:quality_study}
To explore how the quality of pseudo relevance set influences the effectiveness of our approach, we study the correlation between our approach's improvement in percentage over baseline and the quality of pseudo relevance set. The results are presented in Figure \ref{fig:sens_prf}. In Figure \ref{fig:sens_prf}, the x-axis represents the quality of the pseudo relevance set, measured by $AP@k$, while the y-axis represents our approach's improvement in percentage over baseline. Each sample point in Figure \ref{fig:sens_prf} represents a query and $k$ is the optimized number of feedback documents parameter for training queries. In addition, Table \ref{tab:ap_k} presents the mean of our approach's improvement in percentage over baseline at different levels of pseudo relevance set quality. Only the results obtained by $BM25+SEM_{d_{pv}}$ are presented in this section as it has overall better performance than $BM25+SEM_{d_{add}}$ in our experiments in previous sections. From Figure \ref{fig:sens_prf} we can see that the performance of our approach does not obviously correlate to the quality of pseudo relevance document set. According to Table \ref{tab:ap_k}, our approach achieves improvement over the baseline on average at different levels of pseudo relevance set quality for all collections with only one exception on GOV2, which has only 5 queries with $AP@k > 0.9$. Moreover, according to Table \ref{tab:ap_k}, our approach appears to be robust with respect to the low-quality pseudo relevance set $D_{PRF}^{k}$. Also, we find no clear pattern on how the effectiveness of our approach is related to the quality of pseudo relevance set, measured by $AP@k$. We then conduct additional experiment on disk1\&2, by replacing the top-20 initial results returned by $BM25_{PRF}$ with 20 randomly selected non-relevant documents from the qrels. The $BM25_{PRF}$ baseline obtains MAP=0.1811 in this case. We then use the top-30 documents as the pseudo relevance set $D_{PRF}^{k}$, which has only a few relevant documents ranked between 21-30 in the initial results. Despite the poor quality of the pseudo relevance set, our method $BM25_{PRF}+SEM_{d_{pv}}$ still obtains MAP=0.2116, i.e. a 16.84\% statistically significant improvement over the BM25 baseline. One possible explanation is that the non-relevant documents in the PRF set have only a random effect on the SEM score such that the retrieval performance is not evidently affected. In this case, the relevant documents in the $D_{PRF}^{k}$ set are still able to improve the results through our D2D similarity approach.

\subsection{Comparison to Other embedding-based Methods}
\label{sec:comparison}
This section compares our approach with the published results in a list of recently proposed neural models. Note that the results of different approaches cited in this section are obtained under different experimental settings. Therefore, the comparison is only provided for reference. 

\begin{table}[!tbh]
	\centering
	\caption{The comparison between our approach and the locally trained QE in \cite{DBLP:conf/acl/0001MC16} on \textit{nDCG@10}. The results of the locally trained method are the best results reported in \cite{DBLP:conf/acl/0001MC16}. The results of our approach are obtained by $BM25+SEM_{d_{pv}}$.}
	\label{tab:compare}
	\begin{tabular}{|c||c|c|c|}
		\hline
		Model & disk1\&2 & disk4\&5 & CW09B \\ \hline \hline
		$Locally \text{-} Trained$ & 0.563 & 0.517 & 0.258\\ \hline \hline
		%$SEM_{d\text{-}D^k_{PRF}}$ &  0.5779  & 0.5261 & 0.2633\\ \hline
		$Our \; Method$ &  0.5779  & 0.5261 & 0.2633\\ \hline
		
	\end{tabular}
\end{table}

\begin{figure*}[tbh]
	\footnotesize
	\centering
	\subfigure[disk4\&5]{\includegraphics[width=0.4\textwidth]
		{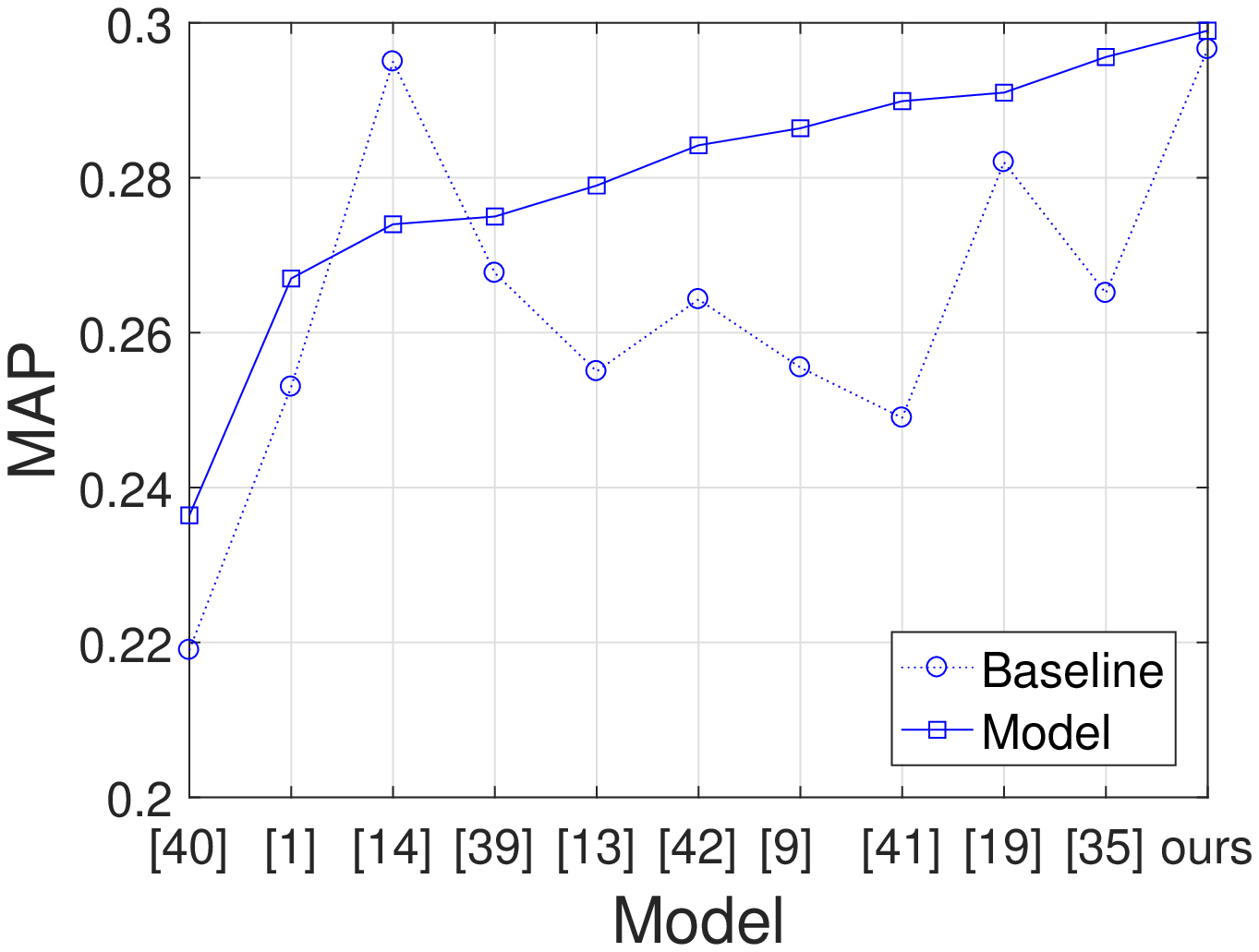} }
	\subfigure[WT10G]{\includegraphics[width=0.4\textwidth]
		{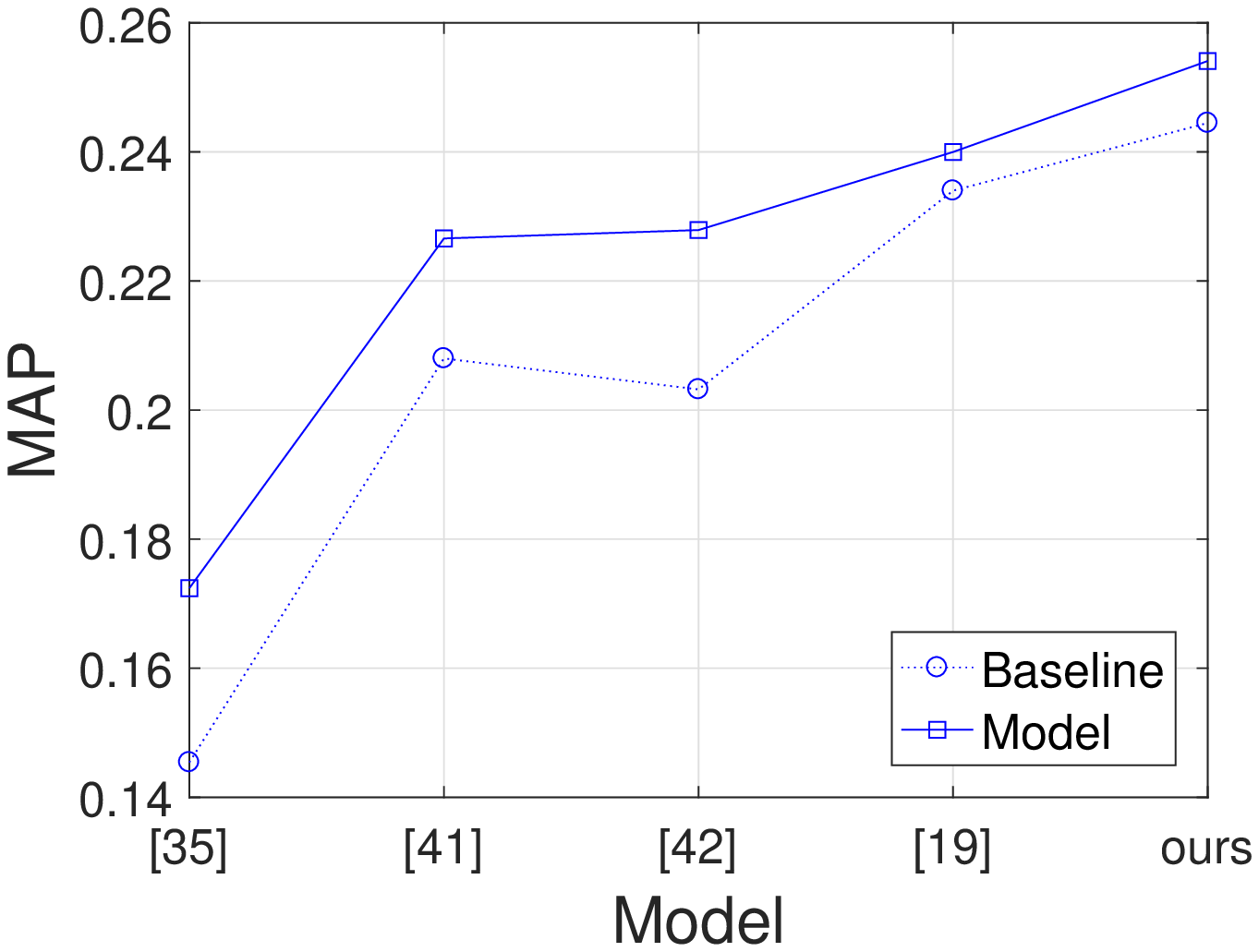} } 
	\subfigure[GOV2]{\includegraphics[width=0.4\textwidth]
		{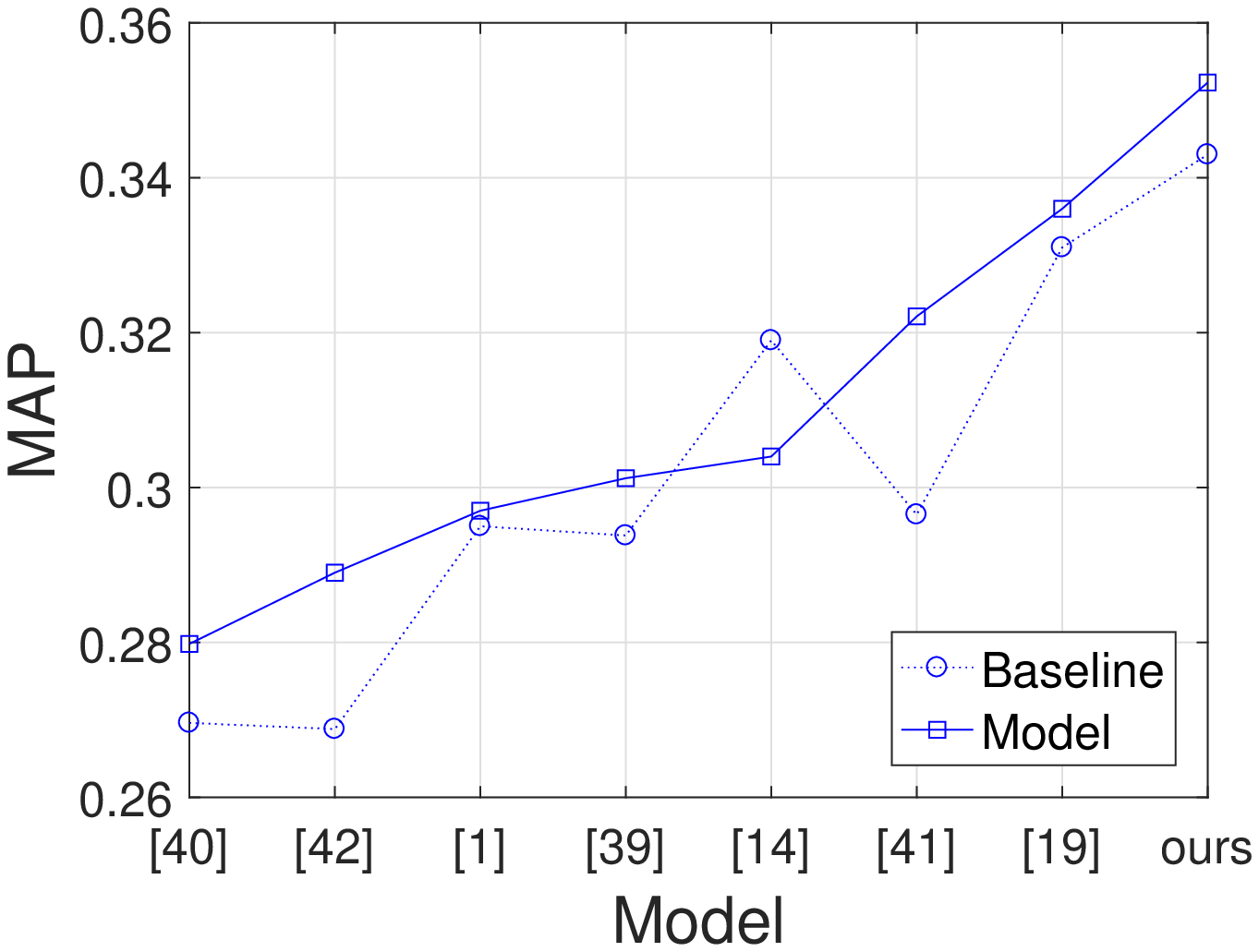} }
	\subfigure[ClueWeb09B]{\includegraphics[width=0.4\textwidth]
		{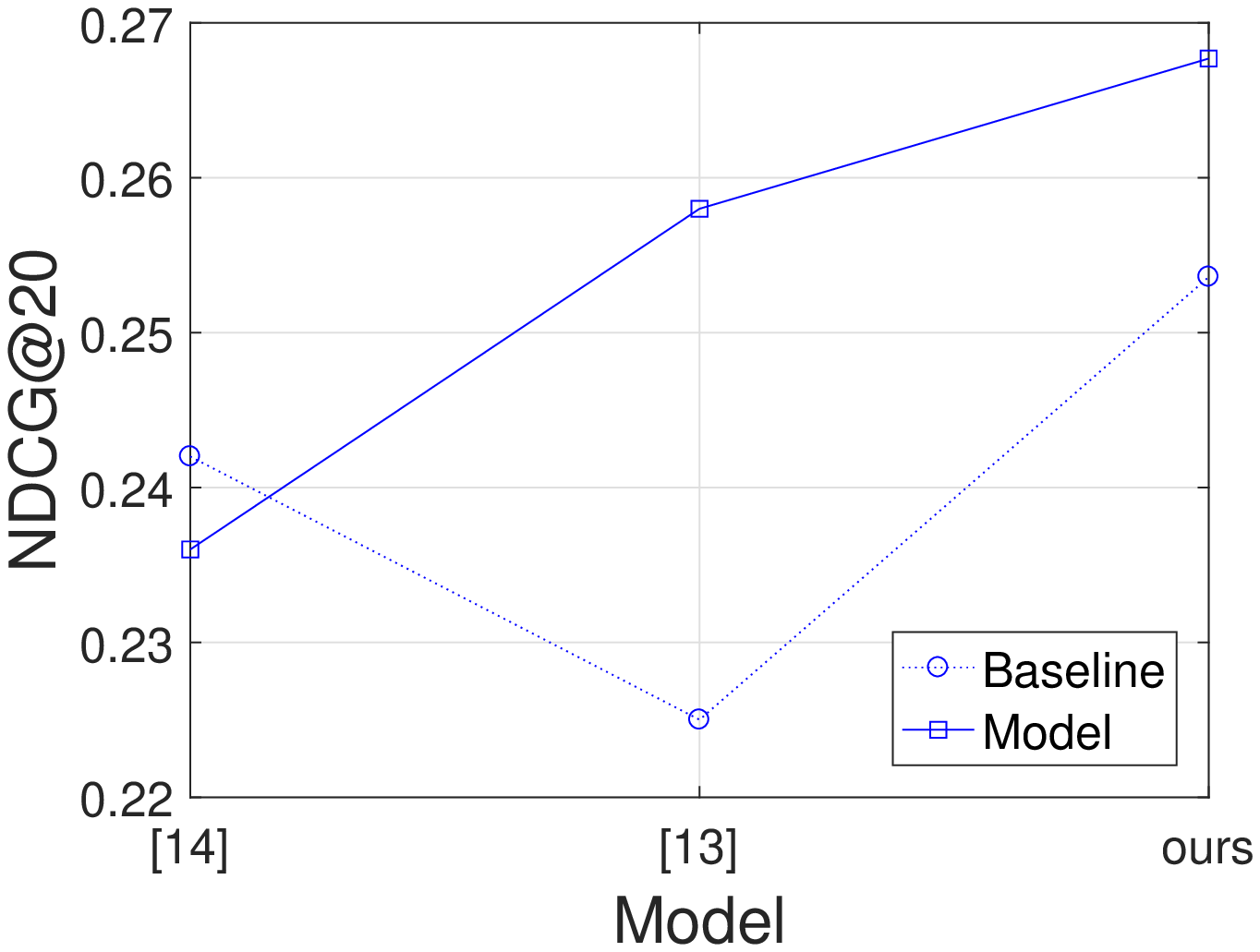} }
	
	\caption{The comparison between our approach and other recently proposed approaches. The results obtained by the baselines and the neural models are both presented.}
	\label{fig:comp_other_works}
\end{figure*}

%In this Section, we compare our approach with two recently proposed state-of-the-art semantics-based models \cite{Tu2016Exploiting,Ai2016Improving}. Table \ref{tab:compare1} presents the comparison between our approach and the results reported in \cite{Tu2016Exploiting}. $ConRank \text{-} BM25$, $ConRank \text{-} LMDir$ and $ConRank \text{-} MATF$ are different applications of the approach in \cite{Tu2016Exploiting}, which are applied based on BM25, QLM and MATF, respectively. According to Table \ref{tab:compare1}, our approach outperforms the approach in \cite{Tu2016Exploiting} on both TREC8 and WT10G test collections in all cases, especially on WT10G. Table \ref{tab:compare2} presents the comparison between our approach and the results reported in \cite{Ai2016Improving}. The proposed \textit{EPV} model in \cite{Ai2016Improving} denotes \textit{enhanced paragraph vector}, and \textit{DRJ} denotes three improvements over the original Distributed Bag of Words version of \textit{Paragraph Vector} (PV-DBOW). \textit{LM} is the language modeling approach to IR which is integrated with \textit{EPV}. From Table \ref{tab:compare2} we can see that our approach has better retrieval performance on both Robust04 and GOV2 test collections. In summary, the comparisons between our approach and the recently proposed state-of-the-art semantics-based retrieval models further demonstrate the effectiveness of our approach.
We first compare the results of the proposed approach with a state-of-the-art query expansion approach based on locally trained embeddings \cite{DBLP:conf/acl/0001MC16}. This approach can also deal with the problem of multiple degrees of similarity by training the word embeddings on only the top-1000 documents. %However, the on-line computational overhead could be an issue in practice since the word embeddings are trained on a per-query basis. 
As only \textit{nDCG@10} is used in \cite{DBLP:conf/acl/0001MC16}, Table \ref{tab:compare} compares the best \textit{nDCG@10} reported in \cite{DBLP:conf/acl/0001MC16} with our approach on each of the three publicly available TREC collections. From the comparison results we can see that our method is at least comparable to \cite{DBLP:conf/acl/0001MC16} on all three TREC test collections. %Therefore, the comparison results demonstrate the effectiveness of our proposed approach.

In addition to \cite{DBLP:conf/acl/0001MC16}, we also compare our approach with other recently proposed works using embeddings or deep neural networks \cite{Ai2016Improving,DBLP:conf/sigir/GangulyRMJ15,DBLP:conf/cikm/GuoFAC16,DBLP:conf/cikm/KuziSK16,DBLP:journals/corr/RoyPMG16,DBLP:conf/ictir/ZamaniC16a,Zamani2016Estimating,Zheng2015Learning,DBLP:conf/cikm/GuoFAC16a}. A recent work \cite{DBLP:conf/cikm/ZamaniDSC16} using matrix factorization for pseudo relevance feedback is also considered. Figure \ref{fig:comp_other_works} compares our proposed D2D similarity approach to a list of recent methods on utilizing embeddings for IR, in which the experiments are also conducted on the publicly available TREC collections. The results obtained by the related methods in Figure \ref{fig:comp_other_works} are taken from those reported in the respective references. The results obtained by our approach are taken from Table \ref{tab:bm25_prf}. The solid lines represent the best results of these works while the dashed lines are the corresponding baselines. We can at least conclude from Figure \ref{fig:comp_other_works} that compared to the previous methods, the proposed D2D similarity approach scores the best retrieval effectiveness over the strongest baseline. In addition, we are puzzled by the fact that our baseline, BM25 with Rocchio's PRF, appears to be stronger than those related studies, although more or less similar models and algorithms are applied. Our guess is that we tune the free parameters, including BM25's parameter $b$, the numbers of feedback documents and expansion terms, and the interpolation parameter $beta$ of Rocchio's PRF, to optimal, while those related studies set the parameters of their baselines to default, or by cross-validation. Nonetheless, we believe it is not a disadvantage in evaluating with as strong as possible baselines. Our baseline can be downloaded from \cite{resultFiles}. Source code implementation is available upon request.

%The results of other works are from their papers's reported results. From Figure \ref{fig:comp_other_works} , we can tell that our approach achieves the best results on all four TREC collections compared to other recently proposed works. This further demonstrate the effectiveness of our approach.

\begin{table}[!tbh]
	\centering
	\caption{The evaluation results on the TREC 2015 CDS task. The difference in percentage is measured against the best result \textit{WSU\text{-}IR} \cite{balaneshinkordan2015wsu} in the task. A statistically significant difference is marked with a *. The best result of each evaluation metric is in \textbf{bold}.}
	\label{tab:cds_2015}
	\begin{tabular}{|c||c|c|}
		\hline
		Method & infNDCG & infAP  \\ \hline \hline
		$BM25_{PRF}$ & 0.2724 & 0.0733  \\ \hline \hline
		\textit{WSU\text{-}IR} & 0.2939 & 0.0842 \\ \hline \hline
		$BM25_{PRF}+SEM_{d\text{-}D^k_{PRF}}$ & 0.2941, +0.07\% & 0.0800, -4.99\%  \\ \hline
		$WSU\text{-}IR+SEM_{d\text{-}D^k_{PRF}}$ & \textbf{0.3111, +5.85\%*} &  \textbf{0.0876, +4.04\%} \\ \hline
		
	\end{tabular}
\end{table}

\vspace{-0.4cm}
\subsection{Application to Clinical Decision Support} \label{sec:cds}

In this section, our proposed approach is applied to the TREC Clinical Decision Support (CDS) task of 2015 \cite{roberts2015overview}, to illustrate the effectiveness of our approach on long documents with long queries. The collection used in the task is composed of 733,138 full-text biomedical articles with an average length of 2,583 tokens, and the topics are patient records with an average length of 80 tokens. The goal of the TREC CDS task is to retrieve relevant biomedical articles with respect to the patient records. The official evaluation metric used in the task is \textit{infNDCG} \cite{roberts2015overview}. The evaluation results are presented in Table \ref{tab:cds_2015}. Note that the semantic relevance score (SEM) is obtained by $\vec{d}_{add}$ or $\vec{d}_{pv}$, denoted as $SEM_{d\text{-}D^k_{PRF}}$. The free parameters in our approach are trained on the topics of the 2014 CDS task. In Table \ref{tab:cds_2015}, \textit{WSU\text{-}IR} \cite{balaneshinkordan2015wsu} is the best run in 2015 \cite{roberts2015overview}, which combines different sources of evidence in a machine learning framework. $WSU\text{-}IR+SEM_{d\text{-}D^k_{PRF}}$ is a linear combination of \textit{WSU\text{-}IR} with our proposed semantic relevance score (SEM). According to the results, a straight-forward application of our approach ($BM25_{PRF}+SEM_{d\text{-}D^k_{PRF}}$) is able to achieve the best official evaluation metric, infNDCG, in this task. In addition, a linear combination of \textit{WSU\text{-}IR} with SEM has statistically significant improvement over \textit{WSU\text{-}IR} in infNDCG, suggesting that it would be beneficial to include semantic relation as an important feature in a learning to rank framework.

\section{Conclusions and Future work}
\label{sec:conclusions}
In this paper, we have studied how to effectively improve the performance of IR models by utilizing the embeddings. In order to overcome the problem ``multiple degrees of similarity'' when using embeddings to IR, we propose a novel D2D-based approach, which measures the semantic similarity between a document and the highly relevant documents, simulated by corresponding pseudo feedback set. Experimental results show that the integration of our proposed semantic relevance score can lead to significant improvements over the classical retrieval models and PRF methods. The effectiveness of different document vector generation models in estimating the semantic similarity between documents, including Word or Para2Vec, LDA, and TF-IDF, are also compared in the experiments. Results show that Word or Para2Vec outperform the other two methods. Finally, it is interesting to see that the effectiveness of our approach does not correlate with the quality of the pseudo feedback set. The D2D similarity approach appears to be robust even if there are only very few not top-ranked relevant documents in the pseudo feedback set, outnumbered by the non-relevant ones. We plan to investigate in an explanation on this finding in future research.

In future research, we also plan to investigate in the application of query reformulation using embeddings in our approach. In addition to the method in \cite{Zamani2016Estimating}, other recent approaches such as those proposed in \cite{Zheng2015Learning,kusner2015word,Rekabsaz2016Uncertainty} are expected to be able to further improve the retrieval performance of our approach. Moreover, we plan to apply our proposed approach to other IR tasks, such as microblog search \cite{lin2014overview}, where the keyword mismatch could have an important impact on the results as microblogs are usually very short.

\begin{acknowledgements}
This work is supported by the National Natural Science Foundation of China (61472391).
\end{acknowledgements}

%\bibliography{ucasir}   % name your BibTeX data base

%\begin{thebibliography}{}
%%
%% and use \bibitem to create references. Consult the Instructions
%% for authors for reference list style.
%\end{thebibliography}

\end{document}